\documentclass[aps,twocolumn,floats,longbibliography,nopacs]{revtex4-1}
\usepackage{graphicx}
\usepackage{dcolumn}
\usepackage{bm}
\usepackage{color}
\usepackage{amsmath}
\usepackage{amssymb}
\usepackage{hyperref}
\usepackage{algorithm}
\usepackage[end]{algpseudocode}

\newcommand{\h}{\mathit{\boldsymbol{h}}}
\newcommand{\X}{\mathit{\boldsymbol{x}}}

\newcommand{\loc}{\mathrm{loc}}

\newcommand{\beq}{\begin{equation}}
\newcommand{\eeq}{\end{equation}}
\newcommand{\ba}{\begin{array}}
\newcommand{\ea}{\end{array}}
\newcommand{\bea}{\begin{eqnarray}}
\newcommand{\eea}{\end{eqnarray}}

\begin{document}

\title{Self-learning projective quantum Monte Carlo simulations guided by restricted Boltzmann machines}

\author{S. Pilati}
\affiliation{School of Science and Technology, Physics Division, Universit{\`a}  di Camerino, 62032 Camerino (MC), Italy}

\author{E. M. Inack}
\affiliation{Perimeter Institute for Theoretical Physics, Waterloo, Ontario N2L 2Y5, Canada}

\author{P. Pieri}
\affiliation{School of Science and Technology, Physics Division, Universit{\`a}  di Camerino, 62032 Camerino (MC), Italy}
\affiliation{INFN, Sezione di Perugia, 06123 Perugia (PG), Italy}


\begin{abstract}
The projective quantum Monte Carlo (PQMC) algorithms are among the most powerful computational techniques to simulate 
the ground state properties of quantum many-body systems. 
However, they are efficient only if a sufficiently accurate trial wave function is used to guide the simulation. In the standard approach, this guiding wave function is obtained in a separate simulation that performs a variational minimization.
Here we show how to perform PQMC simulations guided by an adaptive wave function based on a restricted Boltzmann machine. This adaptive wave function is optimized along the PQMC simulation via unsupervised machine learning, avoiding  the need of a separate variational optimization.
As a byproduct, this technique provides an accurate ansatz for the ground state wave function, which is obtained by minimizing the Kullback-Leibler divergence with respect to the PQMC samples, rather than by minimizing the energy expectation value as in standard variational optimizations.
The high accuracy of this self-learning PQMC technique is demonstrated  for a paradigmatic sign-problem-free model, namely, the ferromagnetic quantum Ising chain, showing  very precise agreement  with the predictions of the Jordan-Wigner theory and of loop quantum Monte Carlo simulations performed in the low-temperature limit.
\end{abstract}

\maketitle

\section{Introduction}
\label{secintro}
The similarity between the imaginary-time Schr\"odinger equation 
and the diffusion equation allows one to simulate quantum many-body systems by stochastically evolving a (typically large) population of 
random walkers~\cite{andersonJB}.
This is the basis of so-called projective quantum Monte Carlo algorithms (PQMC). 
These algorithms are particularly efficient when they simulate the ground state of sign-problem-free Hamiltonians, since the corresponding wave function is real and nonnegative in a suitable basis.
Using fixed-node and released node methods~\cite{reynolds1982fixed,ceperley1984quantum,ceperley1986quantum,doi:10.1063/1.3193710}, PQMC simulations provide accurate predictions 
even when the sign problem occurs, in particular for many-fermion systems, albeit in general at a larger computational cost.
PQMC algorithms have been used to simulate fundamental quantum systems, including the electron gas~\cite{ceperley1980ground}, molecular systems~\cite{reynolds1982fixed}, liquid and solid Helium~\cite{boronat}, electrons in solids~\cite{foulkes2001quantum,booth2013towards}, ultracold gases~\cite{giorgini1999ground}, quantum spin and lattice models~\cite{trivedi1990ground,sorella}, and nuclear matter~\cite{carlson2015quantum}.
However, these algorithms are efficient only if they are provided with a sufficiently accurate trial wave function, which is used to guide the random walkers towards the most relevant regions of the configuration space.
If this guiding wave function is not included, or if it is not sufficiently accurate, the computational cost of a PQMC simulation scales exponentially with the system size~\cite{nemec,boninsegnimoroni,inack2}.
For various relevant systems, the general form of a suitable guiding wave function is provided by some physical theory, and the details can be further tuned via a variational minimization of the energy expectation value.
However, an appropriate theory is not always available, and the variational optimization might turn out to be an extremely challenging computational task. Even more, an inaccurate ansatz for the guiding wave function might lead to biased predictions, in particular when different phases closely compete, as in many strongly-correlated and/or disordered many-body systems.
This problem is especially relevant in the field of adiabatic quantum computing, where quantum Monte Carlo (QMC)  algorithms are being used to simulate how quantum annealers solve complex optimization problems~\cite{santorotheory,boixo2014evidence,inack,boixo2016computational,troyerheim}.
In fact, the Hamiltonians corresponding to typical instances of hard optimization problems can be written in the form of random Ising models. 
These models are characterized by glassy ground states, for which an accurate ansatz is hard to guess~\cite{santoroGFMC}.
%
%
\begin{figure}[h]
\begin{center}
\includegraphics[width=1.0\columnwidth]{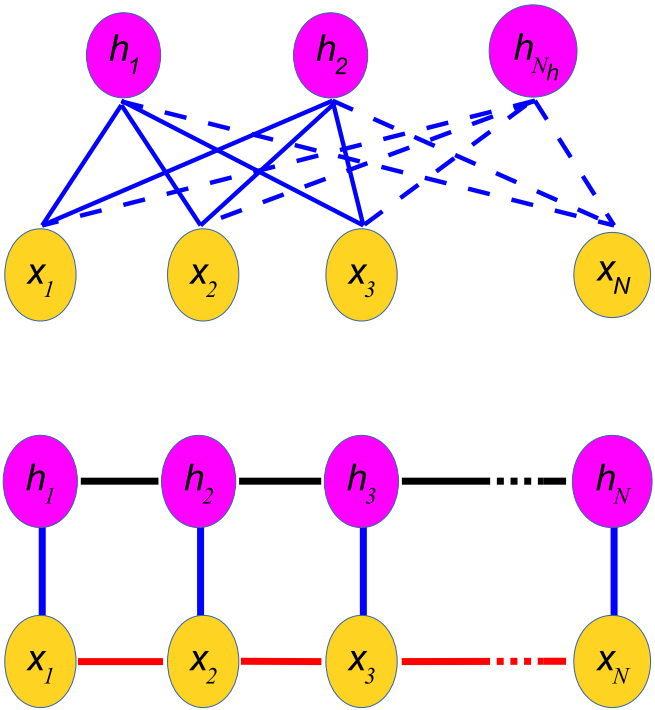}
\caption{(color online). Connectivity structure of the restricted Boltzmann machine (RBM)~\cite{carleotroyer}
employed in this work as PQMC guiding wave function (upper panel) and of the unrestricted Boltzmann machine employed in Ref.~\cite{inack3} (lower panel). 
The latter is analogous to the shadow wave function used for quantum fluids and solids~\cite{reatto1988shadow,shadowvitiello}.
The visible spins are labelled by $x_j$, with $j=1,\dots,N$, while the hidden spins by $h_i$, with $i=1,\dots,N_h$.
The segments indicate the allowed interactions. 
}
\label{fig1}
\end{center}
\end{figure}
%

Wave functions based on a certain type of generative neural network, 
specifically, restricted Boltzmann machines (RBM) [see Fig.~\ref{fig1}, upper panel], 
have been recently proposed as generic variational ansatzes for quantum spin models~\cite{carleotroyer}.
Their representational power, their entanglement properties, and more elaborate versions are currently the subject of very intense research activity~\cite{deng2017quantum,gao2017efficient,saito,chen2018equivalence,glasser2018neural,freitas2018neural,cai2018approximating,saito2,carleo2018constructing,luo2018backflow,kochkov2018variational,PhysRevLett.120.205302,beach2019making,mcbrian2019ground,collura2019descriptive,kessler2019artificial,nagy2019variational}.
They appear to be particularly useful when no other physics-inspired ansatz is available.
In this paper, we show how to employ them as guiding wave functions 
in PQMC simulations.
In a previous article~\cite{inack3}, a neural network state that mimics a type of an unrestricted Boltzmann machine (see Fig.~\ref{fig1}, lower panel)
has been employed as guiding wave function.
In the case of Ref.~\cite{inack3}, the neural network parameters had to be obtained via a separate variational optimization,
and the PQMC algorithm had to be overloaded with the sampling of additional hidden variables.
Here we show that, using unsupervised machine learning algorithms, 
 the RBM can be trained 
along the PQMC simulations directly from random-walker configurations.
One obtains an adaptive guiding wave function that improves during the PQMC simulation, 
allowing one to completely eliminate the bias due to the finite random-walker 
population~\cite{inack2} and to drastically reduce the statistical fluctuations.
As a testbed, we consider a paradigmatic sign-problem free model, 
namely, the ferromagnetic quantum Ising chain, making comparison against 
the exact predictions of the Jordan-Wigner theory and of loop quantum Monte Carlo (QMC) simulations 
performed with the ALPS library~\cite{bauer2011alps} in the low-temperature limit. Excellent agreement is obtained 
for ground-state energies and for the average magnetization, above, below, and at 
the ferromagnetic quantum critical point.
A (possibly useful) byproduct of the self-learning procedure described here is 
an optimized RBM ansatz. This ansatz is developed by minimizing a Kullback-Leibler divergence, rather than the common minimization of the variational energy~\cite{torlaitomography}. 
In the model considered here, the self-learned RBM wave function turns out to be comparatively as
accurate as the one obtained via variational minimization using a sophisticated optimization method such as 
the stochastic reconfiguration algorithm~\cite{PhysRevLett.80.4558} as implemented in the NetKet library~\cite{carleo2019netket}.

The paper is organized as follows.
Section~\ref{secmethod} introduces the model Hamiltonian we address and describes 
the PQMC algorithm, the RBM variational wave-function, as well as the adaptive unsupervised 
learning procedure.
Numerical results and comparison with previous theories are presented in Section~\ref{secresults}.
Section~\ref{secconclusions} summarizes our findings and discusses possible future extensions.

\section{Method}
\label{secmethod}
As a test bed for the self-learning PQMC simulations we consider the one-dimensional ferromagnetic quantum Ising Hamiltonian, defined as:
\begin{equation}
\hat{H}=-J\sum_{j=1}^{N-1} {\sigma}^{z}_{j} {\sigma}^{z}_{j+1} -\Gamma \sum_{j=1}^{N} {\sigma}^{x}_{j} - h \sum_{j=1}^{N} {\sigma}^{z}_{j},
\label{H}
\end{equation}
where $\sigma^x_j$ and $\sigma^z_j$ indicate conventional Pauli matrices acting on the spin at the lattice site $j=1,\dots,N$. $N$ is the total number of spins, and we consider open boundary conditions. The parameter $J>0$ fixes the strength of the ferromagnetic interactions between nearest-neighbor spins.  In the following, we use $J=1$ as unit of the energy scale. 
$\Gamma$ and $h$ are the intensities of the transverse and longitudinal magnetic field, respectively.
Below,  the eigenstates of the Pauli matrix ${\sigma^z_j}$ are denoted as $\left| x_j \right>$. The eigenvalue is $x_j=1$ when $\left|x_j\right>=\left|\uparrow \right>$ and $x_j=-1$ when $\left|x_j \right>=\left|\downarrow \right>$. The quantum states of $N$ spins $\left|\X \right> = \left| x_1 x_2 ... x_N\right>$, with $\X=(x_1,\dots,x_N)$, form the  computational basis considered in the present work. We denote as $\left| \psi \right>$ the quantum state corresponding to the wave-function $\left< x \right| \left. \psi \right> = \psi(\X)$.
\subsection{Projective Quantum Monte Carlo simulations}
The PQMC algorithms allow one to simulate the ground-state of a generic Hamiltonian by stochastically evolving the imaginary-time Schr\"odinger equation. 
The algorithm's accuracy and efficiency greatly improve if one introduces  a suitable ansatz for the ground-state wave-function --- usually called guiding (or trial) wave function and indicated as $\psi_G({\X})$ --- and let the product  $f({\X},\tau)=\psi ({\X},\tau) \psi_G({\X})$ evolve according to the modified imaginary-time Schr\"odinger equation.
This equation reads:
\begin{equation}
\label{masterf}
f(\X,\tau+\Delta \tau) = \sum_{\X^\prime} \tilde{G}(\X,\X^\prime,\Delta \tau) f(\X^\prime,\tau). 
\end{equation}
Here, $\tilde{G}({\X},{\X}^\prime,\Delta \tau)= G(\X,\X^\prime,\Delta \tau)\frac{\psi_G({\X})}{\psi_G({\X^\prime})}$, 
where $G({\X},{\X}^\prime,\Delta \tau)= \left< \X \left| \exp\left[{-\Delta \tau(\hat{H}-E_{\mathrm{ref}})}\right] \right|\X^\prime \right>$ 
is the imaginary-time Green's function for a (short) time step $\Delta \tau$ (we set $\hbar =1$ throughout this paper).
$E_{\mathrm{ref}}$ is a reference energy introduced to stabilize the numerics, as explained below.

We employ the continuous-time algorithm of Refs.~\cite{becca_sorella,SorellaCTGFMC}, which allows one to stochastically simulate the exact (modified) imaginary-time Green's function  $\tilde{G}({\X},{\X}^\prime,\Delta \tau)$,
 avoiding any finite time-step error which would occur with the use of Trotter approximations.
 This is achieved by using the formal zero-time limit of the Green's function and by appropriately sampling the time interval $\delta \tau$ between consecutive single-spin flips. More flips might occur within a time-step $\Delta \tau$.
Since the modified imaginary-time Green's function does not define a standard Markov process, i.e., one driven by a column normalized transition matrix, the simulation has to be performed by stochastically evolving a (large) population of random walkers; these walkers are subjected to updates in configuration space and to a branching process in which they are annihilated or replicated.
The spin-configuration updates $\X^\prime \rightarrow \X$ (with $\X^\prime \neq \X$) are randomly selected from a probability distribution
 proportional to the modified imaginary-time Green's function. The guiding wave function favors updates towards relevant regions of the configuration space.
In the branching process, the random walkers are annihilated or replicated according to the weight factor $w_{\X'} = \exp \left[{-\delta \tau (E_{\loc}(\X') -E_{\mathrm{ref}} )}\right]$, where the local energy is $E_{\loc}(\X')=\sum_{\X} H_{\X',\X}\frac{\psi_G({\X})}{\psi_G({\X^\prime})} $
\footnote{One has to accumulate the weight factors of the updates occurring within the time-step $\Delta \tau$. See Refs.~\cite{SorellaCTGFMC,becca_sorella,inack3} for more details.}.
This weight accounts for the column normalization of the transition matrix.
By dynamically tuning $E_{\mathrm{ref}}$, the size of the random-walker population can be kept very close to a desired target value $N_{\mathrm{w}}$.
To implement the branching process and for the tuning of $E_{\mathrm{ref}}$ we use the textbook recipe of Ref.~\cite{thijssen}.

The continuous-time PQMC algorithm sketched here is more exhaustively described in Refs.~\cite{becca_sorella,SorellaCTGFMC,inack3}, and we refer the interested readers 
to those references for additional implementation details.
The projection can reach long imaginary times $\tau=N_{\mathrm{MC}}\Delta\tau$ by iterating a large number $N_{\mathrm{MC}}$ of  Monte Carlo steps, each step corresponding to a short time step $\Delta\tau$.
In the long imaginary-time limit, the walkers sample spin configurations with a probability distribution proportional to
$f(\X,\tau \rightarrow \infty)=\psi_{0} ({\X}) \psi_G({\X})$ (if $N_w$ is large enough, as explained below), where $\psi_{0} ({\X})$ is the ground-state wave function.
Unbiased predictions of ground-state energies are obtained 
via Monte Carlo integration of the sum 
\beq
E_{\tau\rightarrow\infty} \equiv \frac{\sum_{\X} f\left(\X,\tau\rightarrow\infty\right)E_{\mathrm{loc}}(\X)}{ \sum_{\X} f\left(\X,\tau\rightarrow\infty\right)}.
\eeq
Unbiased prediction of ground-state expectation values of other observables that commute with the Hamiltonian are obtained with analogous formulas. However, for operators $\hat{O}$ that do not commute with the Hamiltonian, the analogous formulas would predict so-called mixed estimators, namely $\left< \psi_{0}\left| \hat{O} \right|  \psi_{G}\right>/\left< \psi_{0}\right| \left.  \psi_{G}\right>$. In general, these are affected by a bias due to the guiding wave function, unless the latter coincides with the ground-state wave function. Nevertheless, if the operator $\hat{O}$ is diagonal in the chosen computational basis, the pure estimator corresponding to $\left< \psi_{0}\left| \hat{O} \right|  \psi_{0}\right>/\left< \psi_{0}\left| \right.  \psi_{0}\right>$ can be determined via the standard forward-walking technique (see, e.g., Ref.~\cite{boronat}). An accurate guiding wave function reduces the computational cost of the forward walking technique.
In the large random-walker population limit, $N_{\mathrm{w}}\rightarrow \infty$, the above Monte Carlo estimates are unbiased, being affected only by statistical fluctuations, which  can be systematically reduced. 
For finite $N_\mathrm{w}$, a systematic bias might arise due to the spurious correlations among replicated walkers~\cite{hetherington1984observations,golinelli1992haldane,nemec,boninsegnimoroni,inack2,pollet2018stochastic}. In Ref.~\cite{inack2}, it was indeed shown that if one does not introduce a guiding wave function, which corresponds to setting $\psi_G({\X})=1$ in the equations above, the random-walker population required to keep this systematic bias below a chosen (small) threshold increases exponentially with the system size. This, 
in turn,  implies a computational cost which scales exponentially.
Instead, if the guiding wave function is exact, i.e. if $\psi_G({\X})=\psi_0({\X})$, the local energy $E_{\mathrm{loc}}(\X)$ is a constant function. This freezes the annihilations and replications of the random walkers in the branching process, eliminating any potential bias.
If $\psi_G({\X})$ is a reasonable approximation for $\psi_0({\X})$, the fluctuations of the random-walker number are anyway reduced compared to the case of the simple PQMC algorithm performed with $\psi_G({\X})=1$,  giving a much faster convergence to the exact $N_w\rightarrow \infty$ limit.\\
\subsection{Boltzmann machines for PQMC algorithms}
In Ref.~\cite{inack3}, it has been shown that a guiding wave function based on an unrestricted Boltzmann machine is sufficiently accurate to drastically reduce the required random-walker population, leading to a polynomially scaling computational cost, at least for the ferromagnetic quantum Ising chain. 
Remarkably, this has been achieved by optimizing just  three variational parameters. The small number of variational parameters to be optimized is the main benefit of the unrestricted architecture of Refs.~\cite{inack3,collura2019descriptive}.
However, using the unrestricted Boltzmann machine requires to sample additional hidden spins, both during the variational optimization
\footnote{In one and in two dimensional models with short-range interactions, the unrestricted Boltzmann machine ansatzes can be mapped to constrained matrix product states and to constrained tensor network states, respectively~\cite{collura2019descriptive}. In one dimension this mapping allows one to implement alternative variational minimization techniques, which avoid the sampling over hidden spins.}
 and, chiefly, during the  PQMC simulations. In fact, the PQMC algorithm has to be extended, combining visible-spin and hidden-spin sampling as explained in Refs.~\cite{whitlock91,inack3}.
An inefficient sampling of the additional hidden spins might introduce further correlations among replicated walkers, leading to a larger finite-$N_w$ bias. This deleterious effect has indeed been observed in PQMC simulation performed close to the ferromagnetic quantum critical point~\cite{inack3}, where the statistical correlations among subsequent hidden-spin configurations are larger. This effect did not lead to an exponentially scaling computational cost in the ferromagnetic quantum Ising chain, but it might be more detrimental for more challenging Hamiltonians.

Below, we show that a guiding wave function based on an RBM allows one to eliminate the finite-$N_{\mathrm{w}}$ bias. 
The additional benefit the RBM ansatz provides is that the optimization of the guiding function can be performed via unsupervised machine learning directly from the random-walker population, with no need for a separate variational optimization.
Furthermore, as opposed to the unrestricted Boltzmann machines, the RBM ansatz does not require to sample additional classical hidden-spin variables. This allows one to avoid overloading the PQMC algorithm with the Monte Carlo updates for the hidden spins.
We argue that these two benefits might be particularly relevant in two setups: in simulation of disordered models, where frustration effects might increase the correlations among successive hidden spin configurations, and in quantum annealing simulations, where the Hamiltonian varies in time and an automatically adaptive guiding wave function would allow one to perform
more efficient simulations of the quantum annealers' dynamics.

Boltzmann machines are generative stochastic neural networks often used to approximate the probability distribution corresponding to  a given population of stochastic samples.
Beyond the visible  spin variables $\X$, they include $N_h$ additional hidden spin variables $h_i=\pm 1$, with $i=1,\dots,N_h$, collectively denoted in the following as $\h=(h_1,\dots,h_{N_h})$.
The probability distribution is written in the form of the Boltzmann weight $P(\X,\h)$ corresponding to a classical Ising Hamiltonian. 
In the case of the RBM, this classical Hamiltonian function reads: 
\beq
\label{ham_rbm}
H_{{\rm RBM}} \left(\X,\h\right)=-\sum_{i,j} J_{ij}h_i x_j -\sum_j a_j x_j - \sum_i b_i h_i.
\eeq
Notice that no visible-visible interactions nor hidden-hidden interactions are allowed. Instead, all visible spins interact with all hidden spins via the coupling parameters $J_{ij}$. One obtains two layers, respectively referred to as visible and hidden layer, with the bipartite all-to-all connectivity illustrated in the upper panel of Fig.~\ref{fig1}.
The parameters $a_j$ and $b_i$, called biases, play the role of longitudinal local magnetic fields. 
%
The couplings $J_{ij}$, together with the bias terms  $a_j$ and $b_i$, define the RBM.
In the following, all these parameters will be collectively denoted as $\bold{W}=\{W_m\}=\{J_{ij}, a_j, b_i\}$, for $i=1,\dots, N_h$ and $j=1,\dots,N$. The index $m$ labels all the RBM parameters.
In practical applications the required number of hidden units is typically $N_h\sim N$. Thus, the number of parameters scales, to leading order, as $N^2$. This quadratic scaling can be reduced to a linear scaling in translationally invariant models~\cite{sohn2012learning,carleotroyer}. However, in view of future studies on disordered systems, we avoid translational invariance adopting open boundary conditions. This allows us to test the unsupervised learning of RBM wave functions in the most generic setup.
In the unrestricted Boltzmann machine considered in Ref.~\cite{inack3}, intra-layer visible-visible and hidden-hidden direct interactions were included, reflecting the same (nearest-neighbor) connectivity of the quantum Hamiltonian in Eq.~\ref{H}. Furthermore, $N_h$ was taken equal to $N$, and inter-layer interactions were allowed only between visible and hidden spins corresponding to the same index, i.e.,  for $i=j$. The resulting connectivity structure is shown in the lower panel of Fig.~\ref{fig1}. This structure is analogous to that of the shadow wave-function used to describe the liquid and solid phases of Helium-4~\cite{reatto1988shadow,shadowvitiello}.

In general, the probability to sample a visible-spin configuration $\X$ is the marginal distribution over all possible hidden-spin configurations $\h$: 
\beq
P( \X ) \equiv \sum_{\h }P(\X,\h) =  \frac{1}{Z}\sum_{\h } \exp\left[-H_{{\rm RBM}} ( \X,\h)\right].
\eeq
Notice that the fictitious temperature and the Boltzmann constant $k_B$ are here equal to unity. The normalization factor is the partition function 
$Z = \sum_{\X,\h} \exp\left[ -H_{{\rm RBM}}  \left(\X,\h\right) \right]$.
Due to the absence of intra-layer interactions in the RBM, the hidden-spin configurations can be analytically traced out, 
resulting in a marginal distribution 
$P(\X) \propto \exp\left(\sum_j a_j x_j\right) \prod_i F_i(\X)$, where $F_i(\X)=2\cosh\left[b_i + \sum_j J_{ij} x_j\right]$.
In Ref.~\cite{carleotroyer}, it has been proposed to use the function $P(\X)$ to define an (unnormalized) ground-state wave-function.
In order to describe both amplitude and phase, the RBM parameters $\bf{W}$ should, in general, be complex valued.
However,  we consider here models whose ground-state wave function can be assumed to be real and nonnegative in a suitable basis; 
therefore, the RBM parameters can be restricted to have real parameters.
Extensions to complex-valued ground-states for, e.g., fermionic systems, have recently been addressed~\cite{luo2018backflow,choo2019study,2019arXiv190600463F}.
In Ref.~\cite{carleotroyer}, the RBM parameters $\bf{W}$ were determined via variational minimization of the expectation value 
$  \left< \psi_{{\rm RBM}} \left| \hat{H} \right|  \psi_{{\rm RBM}} \right>/\left< \psi_{{\rm RBM}} \left| \right.  \psi_{{\rm RBM}} \right>$, where $\left| \psi_{{\rm RBM}}  \right>$ indicates the quantum state corresponding to the (unnormalized)
wave-function $\left<\X | \psi_{{\rm RBM}}  \right> \propto P(\X)$.
The  optimization of the variational parameters was performed using the stochastic reconfiguration algorithm~\cite{SorellaCTGFMC}. 
In the following, the corresponding minimal variational energy, obtained using the Netket~\cite{carleo2019netket} library, will be indicated as $E_{\mathrm{var. min.}} $.
In Ref.~\cite{carleotroyer}, the variational-minimization approach has been referred to as reinforcement learning, due to the close similarity with the reinforcement-learning techniques used in the field of machine learning.
However, in typical machine learning applications of RBMs the parameters $\bf{W}$  are determined via unsupervised machine-learning techniques~\cite{ackley1985learning},
trying to learn the (unknown) distribution corresponding to a (typically large) dataset of stochastic samples. 
The unsupervised learning approach has already been used in Ref.~\cite{torlaitomography} to perform quantum states tomography.
Here, we adopt it to extract reasonably  accurate variational ansatzes from the random-walker population generated by the PQMC algorithm. As discussed below,  these ansatzes can then be used in an adaptive scheme as guiding wave functions to boost the efficiency of the PQMC simulation itself. 
On the one hand, this scheme allows one to eliminate the bias originating from the finite $N_w$. On the other hand, it allows one to eliminate the residual error in the RBM ansatz, providing unbiased predictions of ground-state properties.

\subsection{Unsupervised learning of adaptive RBM guiding wave functions}
In the unsupervised learning approach, the RBM parameters are determined by maximizing the log-likelihood of a (typically large) training set: 
$L(\bold{W}) = \sum_w \ln P(\X_w)$, where $w$ labels the instances in the training set. It can be shown that this corresponds to the minimization of the  so-called 
Kullback-Leibler divergence (see, e.g.,~\cite{fischer2012introduction}). In general, the Kullback-Leibler divergence between two distributions $p(\X)$ and $q(\X)$ is defined as:
\begin{equation}
\mathrm{KL} \left( q \left| \right| p\right) = \sum_{\X} q(\X) \ln\left(q(\X) /p(\X) \right).
\end{equation}
It represents a measure of the distance between two distributions (non-symmetric with respect to the exchange $q \leftrightarrow p$). In the case discussed here, $q(\X)$ is identified with the distribution of the random walkers $f(\X,\tau\rightarrow \infty)$ obtained via PQMC simulations, while $p(\X)$ with the RBM marginal distribution $P(\X)$.
The optimization of the RBM parameters $\bf{W}$  can be performed using the gradient ascent algorithm. It consists in performing  iterative updates  starting from 
an initial (random) guess $\bold{W}^0$. At the step $n=0,1,\dots,N_{\mathrm{steps}}$, one applies the rule:
$W_m^{n+1}=W_m^n + \eta \frac{\partial}{\partial W_m} L(\bold{W}^n) $, where the coefficient $\eta$ is the learning rate. This plain vanilla rule can be improved in various ways, e.g., by including a momentum term proportional to the update performed at the $n-1$ step, by annealing the learning rate, by adopting the  adaptive gradient algorithm (AdaGrad)~\cite{duchi2011adaptive} or the adaptive moment estimation algorithm (Adam)~\cite{kingma2014adam}. 
%
%
For all results reported in this manuscript, the plain vanilla rule is augmented only by adding a momentum term corresponding to $\nu \frac{\partial}{\partial W_m} L(\bold{W}^{n-1}) $, with $\nu$ a rate tuned as discussed below.
Also the learning-rate annealing is adopted, as explained in the following.
%
%
%

The main task in the implementation of the gradient ascent algorithm is the computation of the gradients of the log-likelihood. It is common to use stochastic estimates computed on mini-batches 
of $N_b$ instances  (typically, $N_b\sim10-100$) randomly sampled from the (much larger) training set. 
The formula for the gradients with respect to the couplings $J_{ij}$ is usually stated in the following from:
\begin{equation}
\label{eqgradient}
\frac{\partial L(\bold{W})}{\partial J_{ij}} \propto \left< x_j h_i \right>_{\mathrm{data}} - \left< x_j h_i \right>_{\mathrm{model}}.
\end{equation}

The first term on the right-hand size indicates the average obtained when  the visible 
units are clamped to the data in the mini-batch. 
To determine its value, one uses the probability distribution of the hidden spins $h_i$. Due to the bipartite connectivity, it depends only on the visible-spin configurations.
One has $h_i=1$ with probability $p_{h_i=1}(\X) = 1/ \left[ 1 + \exp\left( -2\sum_j x_j J_{ij} -2b_i \right) \right]$, and  $h_i=-1$ with probability $1-p_{h_i=1}(\X)$.

Determining the second term on the right-hand side of eq.~(\ref{eqgradient}) is more challenging. 
It represents the average obtained when both visible and hidden variables are sampled according to the probability distribution $P(\X,\h)$ defined by the RBM model. 
In machine learning jargon, this term is often referred to as the dreaming phase of the learning process, in contrast to the first term which would correspond to the awake phase.
The average can be determined via Monte Carlo estimation, starting from the visible variables corresponding to the mini-batch instances, and then alternating Gibbs sampling of all hidden variables with fixed visible variables, followed by the sampling of all visible variables with hidden variables fixed at the previously sampled values. The probability to sample $x_j=1$ is analogous to the  formula for the  hidden spins: $p_{x_j=1}(\h) = 1/ \left[ 1 + \exp\left( -2\sum_i h_j J_{ij} -2a_j \right) \right]$.

In principle, this alternated Gibbs sampling should be iterated till the Markov chain equilibrates. This would provide an unbiased estimate of the average.
In practice, repeating a sufficient number of iterations to guarantee that equilibrium has been achieved is often computationally overwhelming. 
The algorithm corresponding to a finite number $k$ of iterations is referred to as 
$k$-step contrastive divergence, since it can be derived by minimizing the difference of two Kullback-Leibler divergences~\cite{hintonConDiv}. 
In the $k\rightarrow\infty$ limit, this algorithm is provably unbiased, meaning that it corresponds to maximizing the log-likelihood of the dataset~\cite{bengio2009justifying}. However, it often turns out to be very accurate also for small $k$.
Formulas for the log-likelihood derivatives with respect to the other RBM parameters, namely the biases $a_j$ and $b_i$, can also be derived, analogously to eq.~(\ref{eqgradient}). Rather than reporting them here, we provide in~\emph{Algorithm}~\ref{alg1} the detailed procedure, which is adapted from Ref.~\cite{fischer2012introduction} to the case of interest to us with binary values $x_j,h_i=\pm1$, rather than the values $1$ and $0$, more common in the machine-learning literature.
The input of the algorithm is a mini-batch of $N_b$ randomly sampled instances, while the output provides the 
 partial derivatives of the log-likelihood $L_{W_m} \equiv   \frac{\partial L(\bold{W})}{\partial W_m}   $. 
 It is convenient to normalize the output, dividing by $N_b$.
For a recent review on the topic, the interested reader is referred to Ref.~\cite{torlai2019review}.
	
\begin{algorithm}[H]
\caption{$k$-step contrastive divergence.} \label{alg1}

\begin{algorithmic}[0]

\State Input: mini-batch $\X_l$, with $l=1,\dots,N_b$
\State Output: partial derivatives $L_{J_{ij}}$, $L_{b_i}$, $L_{a_j}$
\State \hrulefill
\State Initialization: $L_{J_{ij}}=L_{b_i}=L_{a_j}=0$
\State \hrulefill
\For{ $l=1,\dots,N_b$ }

   $\mathbf{x}^0 \leftarrow  \mathbf{x}_l$

   \For{$t= 0,\dots ,k-1$}
       \State set $h_i^t=1$ w. prob. $p_{h_i=1}(\mathbf{x}^t)$, else $h_i^t=-1$
       \State set $x_j^{t+1}=1$ w. prob. $p_{x_j=1}(\mathbf{h}^t)$, else $x_j^{t+1}=-1$
   \EndFor
   \For{$i= 1,\dots ,N_h$ and $j= 1,\dots ,N$ }
      \State $L_{J_{ij}}   \leftarrow L_{J_{ij}} + (2p_{h_i=1}(\mathbf{x}^0)-1)x_j^0 - (2p_{h_i=1} (\mathbf{x}^k)-1)x_j^k$ 
      \State $L_{a_j}    \leftarrow L_{a_{j}} + x_j^0 - x_j^k$ 
      \State $L_{b_i}   \leftarrow L_{b_i} + (2p_{h_i=1}(\mathbf{x}^0)-1) - (2p_{h_i=1} (\mathbf{x}^k)-1)$ 
   \EndFor

\EndFor

\end{algorithmic}
\end{algorithm}

In the adaptive scheme presented in this paper, the above unsupervised learning algorithm is used to train the RBM to describe the (unnormalized) probability distribution $f(\X,\tau\rightarrow\infty)$ corresponding 
to a large random-walker population produced by a PQMC simulation at equilibrium. 
It is in fact known that, in principle, for a sufficiently large $N_h$ an RBM can approximate any discrete distribution (in the sense of the Kullback-Leibler divergence), and many researchers have found that the training algorithms described above are capable of finding optimal, or close to optimal, RBM parameters.

The scheme we propose here involves many consecutive stints, labelled in the following by the index $s=0,1,2,\dots,N_{{\rm stints}}-1$, each including a PQMC simulation with a guiding function $\psi_{G_s}(\X)$ for a long imaginary time $\tau_s$,  followed by a learning stage which is used to construct the new guiding function $\psi_{G_{s+1}}(\X)$. The imaginary time $\tau$ runs up to $\tau_s N_{{\rm stints}}$.
In each learning stage, occurring at the imaginary times $(s+1) \tau_s$, the RBM is trained to describe the equilibrium random-walker distribution produced during the last stint.
For the initial PQMC stint $s=0$, a quite crude guiding wave function $\psi_{G_{s=0}}(\X)$ can be chosen, e.g., a constant function or a wave function based on (the square root of) an RBM with random parameters, $\psi_{G_{s=0}}(\X) \propto \sqrt{P(\X)}$.
After the $s$ stint, the learned RBM distribution is $P_s(\X) \propto \psi_{G_s}(\X)\psi_0(\X)$ (assuming $N_w$ and $\tau_s$ large enough). 
Notice that the learning process converges much faster if in each learning stage the RBM parameters are initialized at the values found in the previous learning stage. 
%

The new guiding function for the stint  $s+1$ is obtained by setting $\psi_{G_{s+1}}(\X)=\sqrt{P_{s}(\X)}$. With this choice the accuracy of the guiding function increases stint after stint.  
Indeed, under the idealized conditions of $N_h$ sufficiently large, successful unsupervised optimization, and $N_w$ and $\tau_s$ large enough, at the $s$ stint the random walkers sample the (unnormalized) distribution 
$f(\X,\tau\rightarrow\infty)=\psi_{G_{s=0}}^{1/2^s}(\X)\psi_{0}^{2-1/2^s}(\X)$.  This indicates an extremely rapid convergence $\psi_{G_{s\rightarrow\infty}}(\X)\rightarrow \psi_0(\X)$.
As a consequence, in less idealized conditions, the (possible) bias due to the finite $N_w$ is expected to disappear after a relatively small number of stints/training stages.
In the next section, we demonstrate that this is indeed the case. This is achieved with an affordable number of hidden spins $N_h$, using a simple implementation of the stochastic gradient ascent algorithm described above.\\

\subsection{Simulation details}
For all results reported in this article, the simulation details are the following. The PQMC simulations are performed with a target number of walkers $N_w=10^4$ and time step $\Delta \tau=0.04$.
 In each stint, the PQMC simulation runs for an imaginary time $\tau_s=20$, which turns out to be sufficiently large to represent the infinite imaginary-time limit $\tau\rightarrow \infty$. 
 A small (randomly selected) fraction of the walker population, namely $N_w/20$, is stored at each PQMC step for the learning stage, excluding the initial time segment of each stint corresponding to $\tau\in\left[s\tau_s,s\tau_s+8\right]$. This avoids correlations and non-equilibrium effects. This protocol provides $N_{\mathrm{train}}\simeq15\times 10^4$ instances for each unsupervised learning stage. The number of stints ranges from $N_{{\rm stints}}=20$ to $N_{{\rm stints}}=50$. This appears to be sufficient to approach the $s\rightarrow \infty$ limit.
 Unsupervised learning is performed after each PQMC stint using the $k$-step contrastive divergence algorithm, performing  a number $N_{\mathrm{steps}}=5\times 10^4$ to $10^5$ of stochastic gradient ascent steps, computed on mini-batches of size $N_b=10$ to $50$. The coefficient of the momentum term is $\nu=\eta/10$ (see description above). The learning rate $\eta$ is kept fixed within each learning stage, but is reduced stage after stage  following the simple empirical annealing protocol $\eta(s)= \eta_0 0.75^s$, where $\eta_0 = 0.01$ is the learning rate at the first learning stage $s=0$. 
 Remarkably, $k=1$ is found to suffice. Tests with $k\approx 30$ provide comparable results. This strongly suggests that in this problem the $k$-step contrastive divergence algorithm is correctly minimizing the Kullback-Leibler divergence. The initial guiding function $\psi_{G_{s=0}}(\X)$ is the square root of an RBM distribution with uniform random couplings $J_{ij}\in \left[-0.025:0.025\right]$. In  the absence of longitudinal magnetic field, i. e. with $h=0$, the bias terms are initialized to zero, while for finite $h$ they are set to uniform random values $a_j,b_i \in \left[0:0.05\right]$.
 It is worth mentioning that, while the setup described here turns out to be adequate for the problems addressed in the present work, it is plausible that even more efficient simulations could be implemented, e.g., by performing more frequent learning stages, storing larger training sets, or changing the optimization algorithm.
\begin{figure}
\begin{center}
\includegraphics[width=1.0\columnwidth]{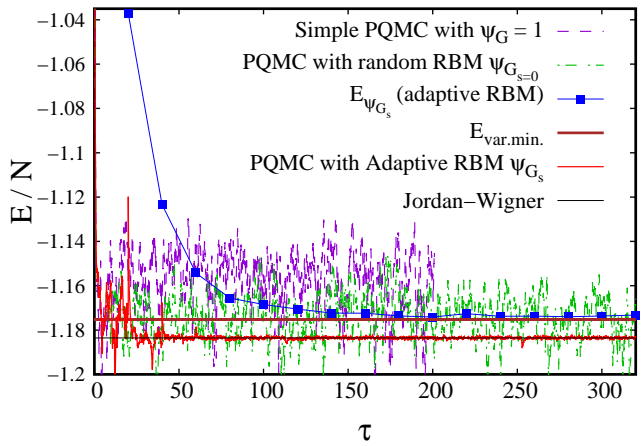}
\caption{(color online). 
Self-learning PQMC simulation of a ferromagnetic quantum Ising chain with $N=80$ spins
 and open boundary conditions.
The transverse-field intensity is $\Gamma=0.85$, the longitudinal-field intensity is $h=0$. 
The energy per spin $E/N$ is plotted as a function of the imaginary time $\tau$.
The (violet) dashed curve indicates the running average in a simple PQMC simulation 
performed without a guiding wave function.
The (dark-green) dot-dashed curve corresponds to a PQMC simulation guided by a fixed 
RBM wave function $\psi_{G_{s=0}}(\X)$ with $N_h=20$ hidden spins and random parameters.
The (red) continuous curve corresponds to the PQMC simulation guided by the adaptive RBM 
wave function $\psi_{G_{s}}(\X)$ trained with unsupervised learning.
The (blue) squares indicate the average energy $E_{\psi_{G_{s}}}$ corresponding to the adaptive RBM wave function $\psi_{G_{s}}(\X)$
obtained after each stint $s$.
The (brown) horizontal bar indicates the minimal variational energy $E_{\mathrm{var. min.}}$ for $N_h=20$ obtained with the NetKet library~\cite{carleo2019netket} using the stochastic reconfiguration algorithm.
The (black) horizontal line corresponds to the exact ground-state energy computed via Jordan-Wigner transformation~\cite{cabrera1987role}.
}
\label{fig2}
\end{center}
\end{figure}
\begin{figure}
\begin{center}
\includegraphics[width=1.0\columnwidth]{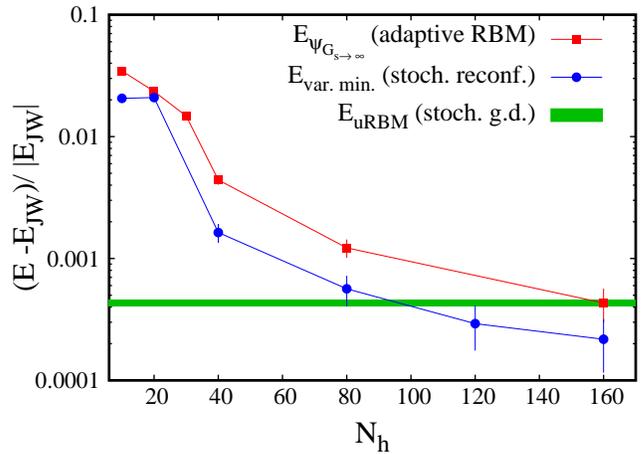}
\caption{(color online). 
Relative error $(E-E_{{\rm JW}} )/\left|E_{{\rm JW}} \right|$ of the estimated ground-state energy $E$ with respect to the (exact) Jordan-Wigner result $E_{{\rm JW}} $, as a function of the  hidden-spin number $N_h$.
The system parameters are: $N=80$, $\Gamma=1$, and $h=0$.
The (red) squares indicate the energy expectation value of the optimal adaptive RBM wave-function $\psi_{G_{s\rightarrow \infty}}(\X)$, optimized 
via unsupervised learning along the PQMC simulation. This optimization corresponds to the minimization of the Kullback-Leibler  divergence.
The (blue) circles correspond to the minimal variational energy for an RBM wave-function obtained with the NetKet library using the stochastic reconfiguration algorithm. 
As a reference, the minimal variational energy corresponding to the unrestricted Boltzmann machine (uRBM) wave function~\cite{inack3} is indicated by the horizontal (green) bar. The width represents (twice) the statistical error. The uRBM wave function is optimized using the stochastic gradient descent algorithm.
}
\label{fig3}
\end{center}
\end{figure}

\begin{figure}[t]
\begin{center}
\includegraphics[width=1.0\columnwidth]{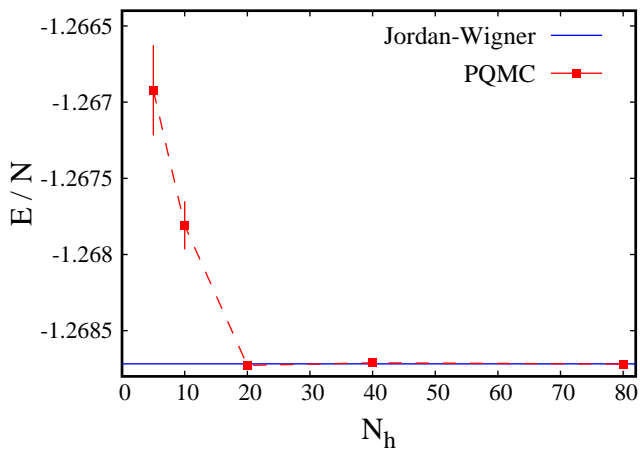}
\caption{(color online). 
Energy per spin $E/N$ obtained from a PQMC simulation guided by 
the optimized adaptive RBM wave functions $\psi_{G_{s \rightarrow \infty}}(\X)$ with different numbers of hidden spins $N_h$.
 The system parameters are: $N=80$, $\Gamma=1$, and $h=0$.
The (blue) horizontal line indicates the exact ground-state energy computed via the Jordan-Wigner (JW) transformation.
}
\label{fig4}
\end{center}
\end{figure}

\section{Results}
\label{secresults}
This section focuses on verifying the accuracy of the self-learning PQMC scheme described in the previous Section. The ferromagnetic quantum Ising chain~(\ref{H}) is used as a testbed.
The PQMC predictions for the ground-state energy are compared against the exact value computed via the Jordan-Wigner transformation~\cite{cabrera1987role}. 
Notice that the comparison is made with the Jordan-Wigner results corresponding to the same system size and the same (open) boundary condition employed in the PQMC simulations.
Let's first focus on a chain with $N=80$ spins, in the absence of a longitudinal field (i.e., with $h=0$), with a transverse field of intensity $\Gamma=0.85$. This value is in the ferromagnetic phase, close to the quantum critical point, which corresponds to $\Gamma=1$ . This is the regime where the systematic bias due to the finite random-walker population is  more sizable~\cite{inack2,inack3}. [For moderately large $N_w$, the bias is larger around $\Gamma \approx 0.85$. However, in the $N_w\rightarrow\infty$ limit, the slowest vanishing of this bias occurs at the quantum critical point $\Gamma=1$, in the thermodynamic limit.]

Indeed, as shown in Fig.~\ref{fig2}, a simple PQMC simulation performed without guiding wave function, i.e., with $\psi_G(\X)=1$, and also one guided by a fixed RBM wave function $\psi_{G_{s=0}}(\X)$ with $N_h=20$ hidden units and random parameters (see previous section for more details), provide energies with a noticeable upward bias of $\sim 1-2\%$. This bias occurs despite the relatively large population of random walkers $N_w=10^4$.
Instead, the self-learning PQMC simulations guided by the adaptive wave function $\psi_{G_s}(\X)$ display a radically different behavior. In the first stint ($s=0$), the guiding wave function $\psi_{G_{s=0}}(\X)$ is the RBM with random parameters described in the previous section; the systematic bias is, again, noticeable. However, this bias completely disappears after a few stints/learning-stages $s\gtrsim3$. Not only the bias disappears, but also the statistical fluctuations are drastically suppressed, leading to an extremely accurate and precise agreement with the Jordan-Wigner prediction.
This indicates that the RBM is capable of accurately learning the random-walker distribution, leading to an adaptive guiding wave function $\psi_{G_{s\rightarrow \infty}}(\X)$ sufficiently similar to the ground-state $\psi_0(\X)$ to eliminate the systematic bias in the PQMC simulation.
To quantify the accuracy of the learned adaptive wave functions $\psi_{G_s}(\X)$, we compute the corresponding energy expectation values 
\beq
E_{\psi_{G_s}}  \equiv  \frac{\langle \psi_{G_s} | \hat{H} |  \psi_{G_s}\rangle} {\langle \psi_{G_s} |  \psi_{G_s} \rangle}.
\eeq
 These computations are performed via Monte Carlo integration using a simple Metropolis sampling algorithm. The corresponding results are displayed in Fig.~\ref{fig2}, placed at the final imaginary-time of the corresponding PQMC stint $\tau=\tau_s(s+1)$.
For small $s$, due to the memory of the initial guiding wave function $\psi_{G_{s=0}}(\X)$ and, to a lesser extent,  to the bias affecting the PQMC algorithm, this expectation value largely deviates from the exact ground-state energy.  This deviation might also originate from limitations in the learning process due to the finite number of learning updates $N_{\mathrm{steps}}$ and/or of the instances in the training set $N_b$. After a few stints/learning-stages the deviation from the ground-state energy significantly reduces, indicating that the adaptive RBM wave function better approximates the ground state.
Remarkably, after a number  $s \gg 1$ of stints and learning stages, $E_{\psi_{G_s}}$ approaches the optimal variational estimate $E_{\mathrm{var. min.}}$ (defined in the previous section) for an RBM wave function obtained using the NetKet library~\cite{carleo2019netket} for the same value of $N_h$. This library implements various sophisticated optimization algorithms. The results reported here are obtained using the stochastic reconfiguration method, which appears to be the most effective algorithm for the model addressed here.
The agreement with the NetKet variational prediction indicates that the unsupervised learning algorithm is capable of identifying an optimal, or close to optimal, ground-state wave function from the random walkers sampled by the PQMC algorithm.

The adaptive RBM wave function $\psi_{G_{s}}(\X)$ with $N_h=20$ is sufficiently accurate to eliminate the bias and to boost the efficiency of the PQMC simulation. However, the corresponding energy expectation value $E_{\psi_{G_{s}}}$ does not precisely agree with the exact ground-state energy, even after a number of stints $s\rightarrow \infty$.
%
%
\begin{figure}
\begin{center}
\includegraphics[width=1.0\columnwidth]{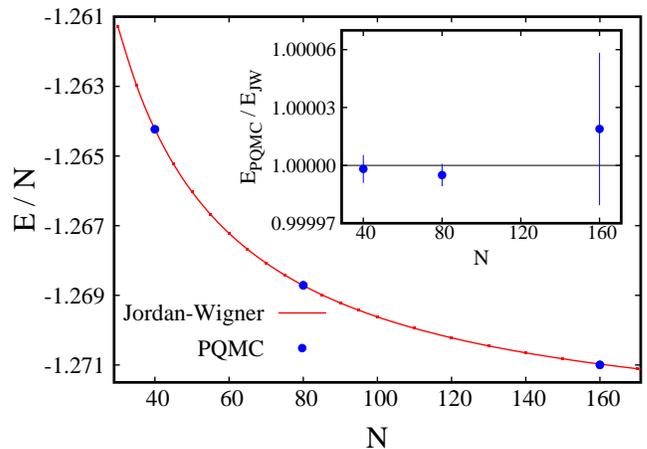}
\caption{(color online). 
Main panel: energy per spin $E/N$ as a function of the system size $N$, at the quantum critical point $\Gamma=1$ with $h=0$. 
The (red) dots and interpolating line indicate the exact Jordan-Wigner result. The (blue) circles indicate the results of PQMC simulation 
guided by the optimized adaptive RBM wave function $\psi_{G_{s\rightarrow\infty}}(\X)$ with $N_h=N/2$ hidden spins.
Inset: ratio of the PQMC result $E_{\mathrm{PQMC}}$  and the exact (Jordan-Wigner) ground-state energy $E_{\mathrm{JW}}$.
}
\label{fig5}
\end{center}
\end{figure}
%
As shown in Fig.~\ref{fig3}, this residual deviation can be systematically reduced by increasing the number of hidden neurons $N_h$ in the RBM. This analysis is shown for the quantum critical point  $\Gamma=1$, where it is more challenging to accurately approximate the ground-state with neural network ansatzes~\cite{carleotroyer,inack3,beach2019making,collura2019descriptive}. The system size is, again, $N=80$.
One also observes that here the NetKet variational estimates $E_{\mathrm{var. min.}}$ for small $N_h$ are slightly lower than $E_{\psi_{G_{s\rightarrow\infty}}}$. This might indicate that the stochastic reconfiguration algorithm is able to better optimize the RBM parameters. However, one should also consider that when a variational ansatz is not very accurate, and the exact ground state is not contained in the manifold defined by the variational parameters, minimizing the energy expectation value does not necessarily provide a better overall description of the ground state. One cannot exclude that the adaptive RBM wave function $\psi_{G_{s\rightarrow\infty}}(\X)$ produced by the unsupervised learning protocol, which aims at minimizing the Kullback-Leibler divergence with respect to the PQMC samples rather than at minimizing the energy expectation value, would provide more accurate predictions of physical properties other than the energy.
As a reference, Fig.~\ref{fig3} displays also the minimal variational energy corresponding to the unrestricted Boltzmann machine ansatz of Ref.~\cite{inack3}. This is obtained by minimizing the energy expectation value with an implementation of the stochastic gradient descent algorithm which samples the additional hidden spin variables. This variational  estimate is comparable to the one corresponding to an RBM wave-function with $N_h=80-100$ hidden spins.
More importantly, if the number of hidden units is too small, the adaptive RBM wave function might not be sufficiently accurate to eliminate the systematic bias of the PQMC simulation, even after many stints/learning-stages. This effect is visualized in Fig.~\ref{fig4}, again for $\Gamma=1$. Indeed, one notices that,  when the adaptive guiding function has $N_h=5-10$ hidden spins, the PQMC results are affected by a small bias $\sim 0.1\%$. This bias completely disappears for $N_h\geqslant 20$.

By exploiting the adaptive RBM wave function, unbiased PQMC results can be obtained also for larger system sizes.  Figure~\ref{fig5} shows data for different system sizes (up to $N=160$). With a number of hidden spins $ N_h=N/2$, one sees from the inset that the relative error with respect to the exact result is as small as $\sim 10^{-3}\%$ even for the largest system size.

\begin{figure}[t]
\begin{center}
\includegraphics[width=1.0\columnwidth]{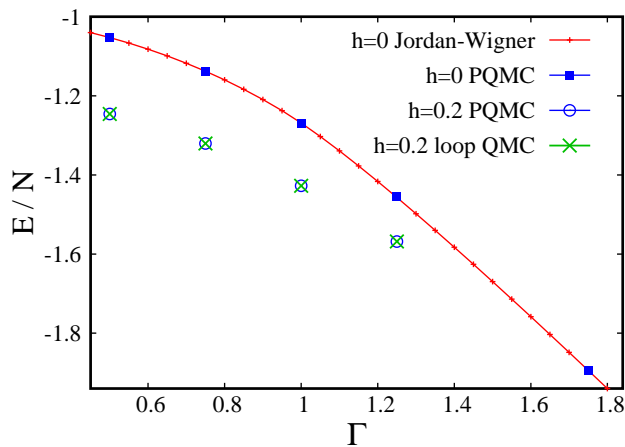}
\caption{(color online). 
Energy per spin $E/N$ as a function of the  transverse magnetic field $\Gamma$, 
for vanishing longitudinal magnetic field, i. e. $h=0$, and for $h=0.2$.
The system size is $N=80$.
At $h=0$, the results of PQMC simulations guided by the optimized adaptive RBM wave function $\psi_{G_{s\rightarrow\infty}}(\X)$ (blue squares) are compared with the Jordan-Wigner theory (red dots and interpolating line).
At $h=0.2$, comparison is made with loop QMC simulations performed at low temperature $ T=0.05$  using the ALPS library~\cite{bauer2011alps}.
}
\label{fig6}
\end{center}
\end{figure}

The PQMC ground-state energy predictions are shown in Fig.~\ref{fig6} as a function of the transverse field $\Gamma$. This plot explores a moderately broad region around the critical point. The system size is $N=80$. Perfect agreement with the Jordan-Wigner prediction is found over the whole region. Results obtained in the presence of a longitudinal field of intensity $h=0.2$ are also shown. 
In this case, the comparison is made against the predictions of loop QMC simulations~\cite{todo2001cluster} performed with the ALPS library~\cite{bauer2011alps}. Notice that the loop QMC algorithm provides finite-temperature results, but we consider here a sufficiently low temperature $T=0.05$ (we set $k_B=1$) such that thermal effects are  negligible.
In the PQMC simulations, the guiding function is, for both $h=0$ and $h=0.2$, an optimized adaptive RBM wave function $\psi_{G_{s\rightarrow\infty}}(\X)$ with $N_h=40$ hidden spins. 
For $h=0$, the bias terms of the RBM are set to zero; in fact, we find that optimizing them does not lead to a more accurate guiding wave function. This is expected, since nonzero bias terms would break the $Z_2$ symmetry of the (finite system) ground state. Instead, for $h=0.2$ the unsupervised learning algorithm leads to sizable bias terms.
%
\begin{figure}
\begin{center}
\includegraphics[width=1.0\columnwidth]{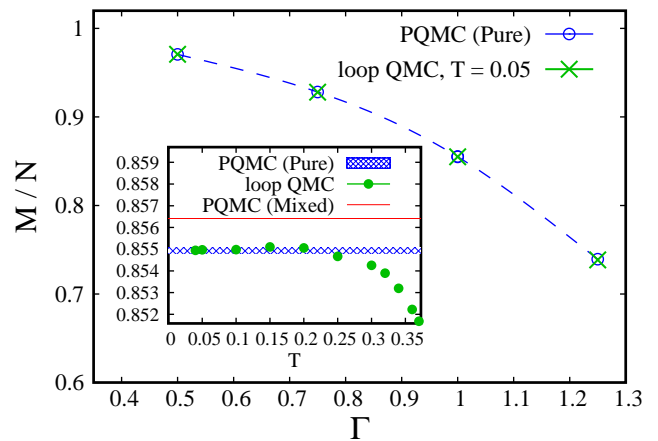}
\caption{(color online). 
Main panel: magnetization per spin $M/N$ as a function of the transverse magnetic field $\Gamma$.
The system size is $N=80$ and the longitudinal-field intensity is $h=0.2$.
The PQMC results obtained with the pure estimator (blue empty circles) are compared with the results of loop QMC simulations (green crosses) 
performed with the ALPS library at the temperature $T=0.05$.
Inset: loop QMC results (green bullets) for $M/N$ as a function of the temperature $T$, for $\Gamma=1$ and $h=0.2$.
The (red) horizontal thin bar corresponds to the zero-temperature PQMC prediction obtained using the mixed estimator.
The horizontal bar with (blue) diagonal pattern corresponds to the pure estimator.
}
\label{fig7}
\end{center}
\end{figure}

The PQMC algorithm provides predictions also for observables other than the energy. In Fig.~\ref{fig7}, the estimates of the ground-state magnetization  
$M \equiv  \left< \psi_0\right| \sum_j \sigma_j^z \left|\psi_0 \right>$, obtained using the forward-walking technique, are shown as a function of the transverse field. A finite longitudinal field $h=0.2$ is included, so that the $Z_2$ symmetry, which would lead to a vanishing average magnetization in finite systems, is broken. The adaptive RBM wave function has $N_h=N/2=40$ hidden units.
 The agreement with the loop QMC predictions is extremely precise. This can be appreciated in the inset of Fig.~\ref{fig7}, where the loop QMC results are plotted as a function of the temperature $T$, showing that thermal effects become indeed irrelevant at the lowest temperatures we consider. The inset displays also the prediction provided by the PQMC mixed estimator (see the definition in the previous section). This is affected by a remarkably small bias $\sim 0.1\%$, indicating that the optimized adaptive RBM guiding function is a good approximation for the ground state. As expected, the pure estimator removes even this small bias.
 
\section{Conclusions}
\label{secconclusions}
Machine learning techniques have been employed in the recent past to implement smart updates in classical and in path-integral Monte Carlo simulations~\cite{huang2017accelerated,PhysRevB.95.241104,PhysRevB.95.041101,PhysRevB.97.205140,li2019accelerating}, leading to a substantial efficiency improvement.
In this paper, we used them to boost the performance of QMC simulations based on projective algorithms.
Specifically, we have introduced a self-learning PQMC algorithm in which the guiding wave function is adaptively developed along the PQMC simulation, avoiding the need of a separate variational optimization. 
This adaptive guiding wave function is based on an RBM, and it is optimized via unsupervised machine learning using the $k$-step contrastive divergence algorithm.
If the number of hidden units is sufficiently large, this adaptive guiding wave function allows one to eliminate any systematic bias in the PQMC simulation and to drastically boost its performance by suppressing the statistical fluctuations.

The self-learning PQMC algorithm appears to be particularly suitable when it is otherwise challenging to guess an accurate ansatz for the ground-state wave function. 
This is the case, e.g., of the disordered Ising Hamiltonians that describe the combinatorial optimization problems commonly addressed in the field of adiabatic quantum optimization.
As a side product of the self-learning PQMC simulations, one obtains an accurate approximation for the ground-state wave function written in the form of an RBM. By construction, this function minimizes the Kullback-Leibler divergence with respect to the random-walker distribution.
We argue that this unsupervised learning approach is complementary to other common techniques for wave function optimization based on the variational minimization of energy expectation values.  Following Ref.~\cite{carleotroyer}, one might refer to these latter techniques as reinforcement learning.
It is indeed plausible that, when the available ansatz is not sufficiently flexible to  exactly describe the ground state, the minimization of the Kullback-Leibler divergence might lead to a fairer description of ground state properties other than the energy.
Our unsupervised learning approach is also complementary to the optimization techniques for neural network states based on supervised machine learning introduced very recently in Ref.~\cite{kochkov2018variational}.

The self-learning PQMC algorithm has been tested in the ferromagnetic quantum Ising chain, obtaining excellent agreement with  the Jordan-Wigner theory and with the loop QMC simulations performed  in the low temperature limit. 
The model considered here is not affected by a sign problem. As a perspective, one can envision the extension of the self-learning PQMC technique to models where the sign problem occurs. Examples are many-fermion models and other non-stoquastic Hamiltonians relevant in adiabatic quantum computing. Fermionic PQMC simulations can be implemented using random walkers that carry a sign.  They have been performed in the past adopting fixed-node and released node approaches, eventually performing annihilation of walkers with opposite sign~\cite{doi:10.1063/1.3193710} to alleviate the suppression of the signal-to-noise ratio associated to the sign problem. In such fermionic simulations, the adaptive RBM guiding function would remain nonnegative, since its aim is to learn the probability distribution of the random walkers. It might allow one to further alleviate  the sign problem by learning the nodal regions, and so reducing the random-walker crossings that lead to the suppression of the signal-to-noise ratio. We leave these endeavors to future investigations. \\

\section*{Acknowledgements}
\noindent
We acknowledge useful discussions with Ferran Mazzanti, Giacomo Torlai, and Giuseppe Santoro. E. M. I. thanks Dustin Lang for his help in performing the simulations on the Perimeter Institute HPC architectures.\\
S. P. and P. P. acknowledge financial support from the FAR2018 project of the University of Camerino and from the Italian MIUR under the project PRIN2017 CEnTraL 20172H2SC4.
S. P. also acknowledges the CINECA award under the ISCRA initiative, for the availability of high performance computing resources and support. Research at Perimeter Institute is supported in part by the Government of Canada through the Department of Innovation, Science and Economic Development Canada and by the Province of Ontario through the Ministry of Economic Development, Job Creation and Trade. This research was supported in part by the National Science Foundation under Grant No. NSF PHY-1748958.



\begin{thebibliography}{73}%
\makeatletter
\providecommand \@ifxundefined [1]{%
 \@ifx{#1\undefined}
}%
\providecommand \@ifnum [1]{%
 \ifnum #1\expandafter \@firstoftwo
 \else \expandafter \@secondoftwo
 \fi
}%
\providecommand \@ifx [1]{%
 \ifx #1\expandafter \@firstoftwo
 \else \expandafter \@secondoftwo
 \fi
}%
\providecommand \natexlab [1]{#1}%
\providecommand \enquote  [1]{``#1''}%
\providecommand \bibnamefont  [1]{#1}%
\providecommand \bibfnamefont [1]{#1}%
\providecommand \citenamefont [1]{#1}%
\providecommand \href@noop [0]{\@secondoftwo}%
\providecommand \href [0]{\begingroup \@sanitize@url \@href}%
\providecommand \@href[1]{\@@startlink{#1}\@@href}%
\providecommand \@@href[1]{\endgroup#1\@@endlink}%
\providecommand \@sanitize@url [0]{\catcode `\\12\catcode `\$12\catcode
  `\&12\catcode `\#12\catcode `\^12\catcode `\_12\catcode `\%12\relax}%
\providecommand \@@startlink[1]{}%
\providecommand \@@endlink[0]{}%
\providecommand \url  [0]{\begingroup\@sanitize@url \@url }%
\providecommand \@url [1]{\endgroup\@href {#1}{\urlprefix }}%
\providecommand \urlprefix  [0]{URL }%
\providecommand \Eprint [0]{\href }%
\providecommand \doibase [0]{http://dx.doi.org/}%
\providecommand \selectlanguage [0]{\@gobble}%
\providecommand \bibinfo  [0]{\@secondoftwo}%
\providecommand \bibfield  [0]{\@secondoftwo}%
\providecommand \translation [1]{[#1]}%
\providecommand \BibitemOpen [0]{}%
\providecommand \bibitemStop [0]{}%
\providecommand \bibitemNoStop [0]{.\EOS\space}%
\providecommand \EOS [0]{\spacefactor3000\relax}%
\providecommand \BibitemShut  [1]{\csname bibitem#1\endcsname}%
\let\auto@bib@innerbib\@empty
\bibitem [{\citenamefont {Anderson}(1975)}]{andersonJB}%
  \BibitemOpen
  \bibfield  {author} {\bibinfo {author} {\bibfnamefont {J.~B.}\ \bibnamefont
  {Anderson}},\ }\bibfield  {title} {\enquote {\bibinfo {title} {{A random-walk
  simulation of the Schr{\"o}dinger equation: H+ 3}},}\ }\href
  {http://aip.scitation.org/doi/abs/10.1063/1.431514} {\bibfield  {journal}
  {\bibinfo  {journal} {J. Chem. Phys.}\ }\textbf {\bibinfo {volume} {63}},\
  \bibinfo {pages} {1499--1503} (\bibinfo {year} {1975})}\BibitemShut {NoStop}%
\bibitem [{\citenamefont {Reynolds}\ \emph {et~al.}(1982)\citenamefont
  {Reynolds}, \citenamefont {Ceperley}, \citenamefont {Alder},\ and\
  \citenamefont {Lester~Jr}}]{reynolds1982fixed}%
  \BibitemOpen
  \bibfield  {author} {\bibinfo {author} {\bibfnamefont {P.~J.}\ \bibnamefont
  {Reynolds}}, \bibinfo {author} {\bibfnamefont {D.~M.}\ \bibnamefont
  {Ceperley}}, \bibinfo {author} {\bibfnamefont {B.~J.}\ \bibnamefont {Alder}},
  \ and\ \bibinfo {author} {\bibfnamefont {W.~A.}\ \bibnamefont {Lester~Jr}},\
  }\bibfield  {title} {\enquote {\bibinfo {title} {{Fixed-node quantum Monte
  Carlo for molecules}},}\ }\href {\doibase 10.1063/1.443766} {\bibfield
  {journal} {\bibinfo  {journal} {J. Chem. Phys.}\ }\textbf {\bibinfo {volume}
  {77}},\ \bibinfo {pages} {5593--5603} (\bibinfo {year} {1982})}\BibitemShut
  {NoStop}%
\bibitem [{\citenamefont {Ceperley}\ and\ \citenamefont
  {Alder}(1984)}]{ceperley1984quantum}%
  \BibitemOpen
  \bibfield  {author} {\bibinfo {author} {\bibfnamefont {D.~M.}\ \bibnamefont
  {Ceperley}}\ and\ \bibinfo {author} {\bibfnamefont {B.}~\bibnamefont
  {Alder}},\ }\bibfield  {title} {\enquote {\bibinfo {title} {{Quantum Monte
  Carlo for molecules: Green’s function and nodal release}},}\ }\href@noop {}
  {\bibfield  {journal} {\bibinfo  {journal} {J. Chem. Phys.}\ }\textbf
  {\bibinfo {volume} {81}},\ \bibinfo {pages} {5833--5844} (\bibinfo {year}
  {1984})}\BibitemShut {NoStop}%
\bibitem [{\citenamefont {Ceperley}\ and\ \citenamefont
  {Alder}(1986)}]{ceperley1986quantum}%
  \BibitemOpen
  \bibfield  {author} {\bibinfo {author} {\bibfnamefont {D.}~\bibnamefont
  {Ceperley}}\ and\ \bibinfo {author} {\bibfnamefont {B.}~\bibnamefont
  {Alder}},\ }\bibfield  {title} {\enquote {\bibinfo {title} {{Quantum Monte
  Carlo}},}\ }\href {\doibase 10.1126/science.231.4738.555} {\bibfield
  {journal} {\bibinfo  {journal} {Science}\ }\textbf {\bibinfo {volume}
  {231}},\ \bibinfo {pages} {555--560} (\bibinfo {year} {1986})}\BibitemShut
  {NoStop}%
\bibitem [{\citenamefont {Booth}\ \emph {et~al.}(2009)\citenamefont {Booth},
  \citenamefont {Thom},\ and\ \citenamefont {Alavi}}]{doi:10.1063/1.3193710}%
  \BibitemOpen
  \bibfield  {author} {\bibinfo {author} {\bibfnamefont {G.~H.}\ \bibnamefont
  {Booth}}, \bibinfo {author} {\bibfnamefont {A.~J.}\ \bibnamefont {Thom}}, \
  and\ \bibinfo {author} {\bibfnamefont {A.}~\bibnamefont {Alavi}},\ }\bibfield
   {title} {\enquote {\bibinfo {title} {{Fermion Monte Carlo without fixed
  nodes: A game of life, death, and annihilation in Slater determinant
  space}},}\ }\href {\doibase 10.1063/1.3193710} {\bibfield  {journal}
  {\bibinfo  {journal} {J. Chem. Phys.}\ }\textbf {\bibinfo {volume} {131}},\
  \bibinfo {pages} {054106} (\bibinfo {year} {2009})}\BibitemShut {NoStop}%
\bibitem [{\citenamefont {Ceperley}\ and\ \citenamefont
  {Alder}(1980)}]{ceperley1980ground}%
  \BibitemOpen
  \bibfield  {author} {\bibinfo {author} {\bibfnamefont {D.~M.}\ \bibnamefont
  {Ceperley}}\ and\ \bibinfo {author} {\bibfnamefont {B.~J.}\ \bibnamefont
  {Alder}},\ }\bibfield  {title} {\enquote {\bibinfo {title} {Ground state of
  the electron gas by a stochastic method},}\ }\href {\doibase
  10.1103/PhysRevLett.45.566} {\bibfield  {journal} {\bibinfo  {journal} {Phys.
  Rev. Lett.}\ }\textbf {\bibinfo {volume} {45}},\ \bibinfo {pages} {566}
  (\bibinfo {year} {1980})}\BibitemShut {NoStop}%
\bibitem [{\citenamefont {Boronat}(2002)}]{boronat}%
  \BibitemOpen
  \bibfield  {author} {\bibinfo {author} {\bibfnamefont {J.}~\bibnamefont
  {Boronat}},\ }\bibfield  {title} {\enquote {\bibinfo {title} {{Monte Carlo
  simulations at zero temperature: helium in one, two, and three
  dimensions}},}\ }in\ \href@noop {} {\emph {\bibinfo {booktitle} {Microscopic
  Approaches to Quantum Liquids in Confined Geometries}}},\ \bibinfo {editor}
  {edited by\ \bibinfo {editor} {\bibfnamefont {E.}~\bibnamefont {Krotscheck}}\
  and\ \bibinfo {editor} {\bibfnamefont {J.}~\bibnamefont {Navarro}}}\
  (\bibinfo  {publisher} {World Scientific, Singapore, 2002},\ \bibinfo {year}
  {2002})\ Chap.~\bibinfo {chapter} {2}, pp.\ \bibinfo {pages}
  {21--90}\BibitemShut {NoStop}%
\bibitem [{\citenamefont {Foulkes}\ \emph {et~al.}(2001)\citenamefont
  {Foulkes}, \citenamefont {Mitas}, \citenamefont {Needs},\ and\ \citenamefont
  {Rajagopal}}]{foulkes2001quantum}%
  \BibitemOpen
  \bibfield  {author} {\bibinfo {author} {\bibfnamefont {W.}~\bibnamefont
  {Foulkes}}, \bibinfo {author} {\bibfnamefont {L.}~\bibnamefont {Mitas}},
  \bibinfo {author} {\bibfnamefont {R.}~\bibnamefont {Needs}}, \ and\ \bibinfo
  {author} {\bibfnamefont {G.}~\bibnamefont {Rajagopal}},\ }\bibfield  {title}
  {\enquote {\bibinfo {title} {{Quantum Monte Carlo simulations of solids}},}\
  }\href {\doibase 10.1103/RevModPhys.73.33} {\bibfield  {journal} {\bibinfo
  {journal} {Rev. Mod. Phys.}\ }\textbf {\bibinfo {volume} {73}},\ \bibinfo
  {pages} {33} (\bibinfo {year} {2001})}\BibitemShut {NoStop}%
\bibitem [{\citenamefont {Booth}\ \emph {et~al.}(2013)\citenamefont {Booth},
  \citenamefont {Gr{\"u}neis}, \citenamefont {Kresse},\ and\ \citenamefont
  {Alavi}}]{booth2013towards}%
  \BibitemOpen
  \bibfield  {author} {\bibinfo {author} {\bibfnamefont {G.~H.}\ \bibnamefont
  {Booth}}, \bibinfo {author} {\bibfnamefont {A.}~\bibnamefont {Gr{\"u}neis}},
  \bibinfo {author} {\bibfnamefont {G.}~\bibnamefont {Kresse}}, \ and\ \bibinfo
  {author} {\bibfnamefont {A.}~\bibnamefont {Alavi}},\ }\bibfield  {title}
  {\enquote {\bibinfo {title} {Towards an exact description of electronic
  wavefunctions in real solids},}\ }\href@noop {} {\bibfield  {journal}
  {\bibinfo  {journal} {Nature}\ }\textbf {\bibinfo {volume} {493}},\ \bibinfo
  {pages} {365} (\bibinfo {year} {2013})}\BibitemShut {NoStop}%
\bibitem [{\citenamefont {Giorgini}\ \emph {et~al.}(1999)\citenamefont
  {Giorgini}, \citenamefont {Boronat},\ and\ \citenamefont
  {Casulleras}}]{giorgini1999ground}%
  \BibitemOpen
  \bibfield  {author} {\bibinfo {author} {\bibfnamefont {S.}~\bibnamefont
  {Giorgini}}, \bibinfo {author} {\bibfnamefont {J.}~\bibnamefont {Boronat}}, \
  and\ \bibinfo {author} {\bibfnamefont {J.}~\bibnamefont {Casulleras}},\
  }\bibfield  {title} {\enquote {\bibinfo {title} {{Ground state of a
  homogeneous Bose gas: A diffusion Monte Carlo calculation}},}\ }\href
  {\doibase 10.1103/PhysRevA.60.5129} {\bibfield  {journal} {\bibinfo
  {journal} {Phys. Rev. A}\ }\textbf {\bibinfo {volume} {60}},\ \bibinfo
  {pages} {5129} (\bibinfo {year} {1999})}\BibitemShut {NoStop}%
\bibitem [{\citenamefont {Trivedi}\ and\ \citenamefont
  {Ceperley}(1990)}]{trivedi1990ground}%
  \BibitemOpen
  \bibfield  {author} {\bibinfo {author} {\bibfnamefont {N.}~\bibnamefont
  {Trivedi}}\ and\ \bibinfo {author} {\bibfnamefont {D.}~\bibnamefont
  {Ceperley}},\ }\bibfield  {title} {\enquote {\bibinfo {title} {{Ground-state
  correlations of quantum antiferromagnets: A Green-function Monte Carlo
  study}},}\ }\href {\doibase 10.1103/PhysRevB.41.4552} {\bibfield  {journal}
  {\bibinfo  {journal} {Phys. Rev. B}\ }\textbf {\bibinfo {volume} {41}},\
  \bibinfo {pages} {4552} (\bibinfo {year} {1990})}\BibitemShut {NoStop}%
\bibitem [{\citenamefont {Buonaura}\ and\ \citenamefont
  {Sorella}(1998)}]{sorella}%
  \BibitemOpen
  \bibfield  {author} {\bibinfo {author} {\bibfnamefont {M.~C.}\ \bibnamefont
  {Buonaura}}\ and\ \bibinfo {author} {\bibfnamefont {S.}~\bibnamefont
  {Sorella}},\ }\bibfield  {title} {\enquote {\bibinfo {title} {{Numerical
  study of the two-dimensional Heisenberg model using a Green function Monte
  Carlo technique with a fixed number of walkers}},}\ }\href
  {https://journals.aps.org/prb/abstract/10.1103/PhysRevB.57.11446} {\bibfield
  {journal} {\bibinfo  {journal} {Phys. Rev. B}\ }\textbf {\bibinfo {volume}
  {57}},\ \bibinfo {pages} {11446} (\bibinfo {year} {1998})}\BibitemShut
  {NoStop}%
\bibitem [{\citenamefont {Carlson}\ \emph {et~al.}(2015)\citenamefont
  {Carlson}, \citenamefont {Gandolfi}, \citenamefont {Pederiva}, \citenamefont
  {Pieper}, \citenamefont {Schiavilla}, \citenamefont {Schmidt},\ and\
  \citenamefont {Wiringa}}]{carlson2015quantum}%
  \BibitemOpen
  \bibfield  {author} {\bibinfo {author} {\bibfnamefont {J.}~\bibnamefont
  {Carlson}}, \bibinfo {author} {\bibfnamefont {S.}~\bibnamefont {Gandolfi}},
  \bibinfo {author} {\bibfnamefont {F.}~\bibnamefont {Pederiva}}, \bibinfo
  {author} {\bibfnamefont {S.~C.}\ \bibnamefont {Pieper}}, \bibinfo {author}
  {\bibfnamefont {R.}~\bibnamefont {Schiavilla}}, \bibinfo {author}
  {\bibfnamefont {K.}~\bibnamefont {Schmidt}}, \ and\ \bibinfo {author}
  {\bibfnamefont {R.~B.}\ \bibnamefont {Wiringa}},\ }\bibfield  {title}
  {\enquote {\bibinfo {title} {{Quantum Monte Carlo methods for nuclear
  physics}},}\ }\href {\doibase 10.1103/RevModPhys.87.1067} {\bibfield
  {journal} {\bibinfo  {journal} {Rev. Mod. Phys.}\ }\textbf {\bibinfo {volume}
  {87}},\ \bibinfo {pages} {1067} (\bibinfo {year} {2015})}\BibitemShut
  {NoStop}%
\bibitem [{\citenamefont {Nemec}(2010)}]{nemec}%
  \BibitemOpen
  \bibfield  {author} {\bibinfo {author} {\bibfnamefont {N.}~\bibnamefont
  {Nemec}},\ }\bibfield  {title} {\enquote {\bibinfo {title} {{Diffusion Monte
  Carlo: Exponential scaling of computational cost for large systems}},}\
  }\href {\doibase 10.1103/PhysRevB.81.035119} {\bibfield  {journal} {\bibinfo
  {journal} {Phys. Rev. B}\ }\textbf {\bibinfo {volume} {81}},\ \bibinfo
  {pages} {035119} (\bibinfo {year} {2010})}\BibitemShut {NoStop}%
\bibitem [{\citenamefont {Boninsegni}\ and\ \citenamefont
  {Moroni}(2012)}]{boninsegnimoroni}%
  \BibitemOpen
  \bibfield  {author} {\bibinfo {author} {\bibfnamefont {M.}~\bibnamefont
  {Boninsegni}}\ and\ \bibinfo {author} {\bibfnamefont {S.}~\bibnamefont
  {Moroni}},\ }\bibfield  {title} {\enquote {\bibinfo {title} {{Population size
  bias in diffusion Monte Carlo}},}\ }\href {\doibase
  10.1103/PhysRevE.86.056712} {\bibfield  {journal} {\bibinfo  {journal} {Phys.
  Rev. E}\ }\textbf {\bibinfo {volume} {86}},\ \bibinfo {pages} {056712}
  (\bibinfo {year} {2012})}\BibitemShut {NoStop}%
\bibitem [{\citenamefont {Inack}\ \emph
  {et~al.}(2018{\natexlab{a}})\citenamefont {Inack}, \citenamefont {Giudici},
  \citenamefont {Parolini}, \citenamefont {Santoro},\ and\ \citenamefont
  {Pilati}}]{inack2}%
  \BibitemOpen
  \bibfield  {author} {\bibinfo {author} {\bibfnamefont {E.}~\bibnamefont
  {Inack}}, \bibinfo {author} {\bibfnamefont {G.}~\bibnamefont {Giudici}},
  \bibinfo {author} {\bibfnamefont {T.}~\bibnamefont {Parolini}}, \bibinfo
  {author} {\bibfnamefont {G.}~\bibnamefont {Santoro}}, \ and\ \bibinfo
  {author} {\bibfnamefont {S.}~\bibnamefont {Pilati}},\ }\bibfield  {title}
  {\enquote {\bibinfo {title} {{Understanding quantum tunneling using diffusion
  Monte Carlo simulations}},}\ }\href {\doibase 10.1103/PhysRevA.97.032307}
  {\bibfield  {journal} {\bibinfo  {journal} {Phys. Rev. A}\ }\textbf {\bibinfo
  {volume} {97}},\ \bibinfo {pages} {032307} (\bibinfo {year}
  {2018}{\natexlab{a}})}\BibitemShut {NoStop}%
\bibitem [{\citenamefont {Santoro}\ \emph {et~al.}(2002)\citenamefont
  {Santoro}, \citenamefont {Marto{\v{n}}{\'a}k}, \citenamefont {Tosatti},\ and\
  \citenamefont {Car}}]{santorotheory}%
  \BibitemOpen
  \bibfield  {author} {\bibinfo {author} {\bibfnamefont {G.~E.}\ \bibnamefont
  {Santoro}}, \bibinfo {author} {\bibfnamefont {R.}~\bibnamefont
  {Marto{\v{n}}{\'a}k}}, \bibinfo {author} {\bibfnamefont {E.}~\bibnamefont
  {Tosatti}}, \ and\ \bibinfo {author} {\bibfnamefont {R.}~\bibnamefont
  {Car}},\ }\bibfield  {title} {\enquote {\bibinfo {title} {Theory of quantum
  annealing of an {Ising} spin glass},}\ }\href
  {http://science.sciencemag.org/content/295/5564/2427.full} {\bibfield
  {journal} {\bibinfo  {journal} {Science}\ }\textbf {\bibinfo {volume}
  {295}},\ \bibinfo {pages} {2427--2430} (\bibinfo {year} {2002})}\BibitemShut
  {NoStop}%
\bibitem [{\citenamefont {Boixo}\ \emph {et~al.}(2014)\citenamefont {Boixo},
  \citenamefont {R{\o}nnow}, \citenamefont {Isakov}, \citenamefont {Wang},
  \citenamefont {Wecker}, \citenamefont {Lidar}, \citenamefont {Martinis},\
  and\ \citenamefont {Troyer}}]{boixo2014evidence}%
  \BibitemOpen
  \bibfield  {author} {\bibinfo {author} {\bibfnamefont {S.}~\bibnamefont
  {Boixo}}, \bibinfo {author} {\bibfnamefont {T.~F.}\ \bibnamefont
  {R{\o}nnow}}, \bibinfo {author} {\bibfnamefont {S.~V.}\ \bibnamefont
  {Isakov}}, \bibinfo {author} {\bibfnamefont {Z.}~\bibnamefont {Wang}},
  \bibinfo {author} {\bibfnamefont {D.}~\bibnamefont {Wecker}}, \bibinfo
  {author} {\bibfnamefont {D.~A.}\ \bibnamefont {Lidar}}, \bibinfo {author}
  {\bibfnamefont {J.~M.}\ \bibnamefont {Martinis}}, \ and\ \bibinfo {author}
  {\bibfnamefont {M.}~\bibnamefont {Troyer}},\ }\bibfield  {title} {\enquote
  {\bibinfo {title} {Evidence for quantum annealing with more than one hundred
  qubits},}\ }\href
  {http://www.nature.com/nphys/journal/v10/n3/full/nphys2900.html} {\bibfield
  {journal} {\bibinfo  {journal} {Nat. Phys.}\ }\textbf {\bibinfo {volume}
  {10}},\ \bibinfo {pages} {218--224} (\bibinfo {year} {2014})}\BibitemShut
  {NoStop}%
\bibitem [{\citenamefont {Inack}\ and\ \citenamefont {Pilati}(2015)}]{inack}%
  \BibitemOpen
  \bibfield  {author} {\bibinfo {author} {\bibfnamefont {E.}~\bibnamefont
  {Inack}}\ and\ \bibinfo {author} {\bibfnamefont {S.}~\bibnamefont {Pilati}},\
  }\bibfield  {title} {\enquote {\bibinfo {title} {Simulated quantum annealing
  of double-well and multiwell potentials},}\ }\href
  {https://link.aps.org/doi/10.1103/PhysRevE.92.053304} {\bibfield  {journal}
  {\bibinfo  {journal} {Phys. Rev. E}\ }\textbf {\bibinfo {volume} {92}},\
  \bibinfo {pages} {053304} (\bibinfo {year} {2015})}\BibitemShut {NoStop}%
\bibitem [{\citenamefont {Boixo}\ \emph {et~al.}(2016)\citenamefont {Boixo},
  \citenamefont {Smelyanskiy}, \citenamefont {Shabani}, \citenamefont {Isakov},
  \citenamefont {Dykman}, \citenamefont {Denchev}, \citenamefont {Amin},
  \citenamefont {Smirnov}, \citenamefont {Mohseni},\ and\ \citenamefont
  {Neven}}]{boixo2016computational}%
  \BibitemOpen
  \bibfield  {author} {\bibinfo {author} {\bibfnamefont {S.}~\bibnamefont
  {Boixo}}, \bibinfo {author} {\bibfnamefont {V.~N.}\ \bibnamefont
  {Smelyanskiy}}, \bibinfo {author} {\bibfnamefont {A.}~\bibnamefont
  {Shabani}}, \bibinfo {author} {\bibfnamefont {S.~V.}\ \bibnamefont {Isakov}},
  \bibinfo {author} {\bibfnamefont {M.}~\bibnamefont {Dykman}}, \bibinfo
  {author} {\bibfnamefont {V.~S.}\ \bibnamefont {Denchev}}, \bibinfo {author}
  {\bibfnamefont {M.~H.}\ \bibnamefont {Amin}}, \bibinfo {author}
  {\bibfnamefont {A.~Y.}\ \bibnamefont {Smirnov}}, \bibinfo {author}
  {\bibfnamefont {M.}~\bibnamefont {Mohseni}}, \ and\ \bibinfo {author}
  {\bibfnamefont {H.}~\bibnamefont {Neven}},\ }\bibfield  {title} {\enquote
  {\bibinfo {title} {Computational multiqubit tunnelling in programmable
  quantum annealers},}\ }\href@noop {} {\bibfield  {journal} {\bibinfo
  {journal} {Nat. Comm.}\ }\textbf {\bibinfo {volume} {7}},\ \bibinfo {pages}
  {10327} (\bibinfo {year} {2016})}\BibitemShut {NoStop}%
\bibitem [{\citenamefont {Heim}\ \emph {et~al.}(2015)\citenamefont {Heim},
  \citenamefont {R{\o}nnow}, \citenamefont {Isakov},\ and\ \citenamefont
  {Troyer}}]{troyerheim}%
  \BibitemOpen
  \bibfield  {author} {\bibinfo {author} {\bibfnamefont {B.}~\bibnamefont
  {Heim}}, \bibinfo {author} {\bibfnamefont {T.~F.}\ \bibnamefont {R{\o}nnow}},
  \bibinfo {author} {\bibfnamefont {S.~V.}\ \bibnamefont {Isakov}}, \ and\
  \bibinfo {author} {\bibfnamefont {M.}~\bibnamefont {Troyer}},\ }\bibfield
  {title} {\enquote {\bibinfo {title} {{Quantum versus classical annealing of
  Ising spin glasses}},}\ }\href
  {http://science.sciencemag.org/content/348/6231/215} {\bibfield  {journal}
  {\bibinfo  {journal} {Science}\ }\textbf {\bibinfo {volume} {348}},\ \bibinfo
  {pages} {215--217} (\bibinfo {year} {2015})}\BibitemShut {NoStop}%
\bibitem [{\citenamefont {Stella}\ and\ \citenamefont
  {Santoro}(2007)}]{santoroGFMC}%
  \BibitemOpen
  \bibfield  {author} {\bibinfo {author} {\bibfnamefont {L.}~\bibnamefont
  {Stella}}\ and\ \bibinfo {author} {\bibfnamefont {G.~E.}\ \bibnamefont
  {Santoro}},\ }\bibfield  {title} {\enquote {\bibinfo {title} {{Quantum
  annealing of an Ising spin-glass by Green’s function Monte Carlo}},}\
  }\href {\doibase 10.1103/PhysRevE.75.036703} {\bibfield  {journal} {\bibinfo
  {journal} {Phys. Rev. E}\ }\textbf {\bibinfo {volume} {75}},\ \bibinfo
  {pages} {036703} (\bibinfo {year} {2007})}\BibitemShut {NoStop}%
\bibitem [{\citenamefont {Carleo}\ and\ \citenamefont
  {Troyer}(2017)}]{carleotroyer}%
  \BibitemOpen
  \bibfield  {author} {\bibinfo {author} {\bibfnamefont {G.}~\bibnamefont
  {Carleo}}\ and\ \bibinfo {author} {\bibfnamefont {M.}~\bibnamefont
  {Troyer}},\ }\bibfield  {title} {\enquote {\bibinfo {title} {Solving the
  quantum many-body problem with artificial neural networks},}\ }\href
  {\doibase 10.1126/science.aag2302} {\bibfield  {journal} {\bibinfo  {journal}
  {Science}\ }\textbf {\bibinfo {volume} {355}},\ \bibinfo {pages} {602--606}
  (\bibinfo {year} {2017})}\BibitemShut {NoStop}%
\bibitem [{\citenamefont {Inack}\ \emph
  {et~al.}(2018{\natexlab{b}})\citenamefont {Inack}, \citenamefont {Santoro},
  \citenamefont {Dell'Anna},\ and\ \citenamefont {Pilati}}]{inack3}%
  \BibitemOpen
  \bibfield  {author} {\bibinfo {author} {\bibfnamefont {E.}~\bibnamefont
  {Inack}}, \bibinfo {author} {\bibfnamefont {G.}~\bibnamefont {Santoro}},
  \bibinfo {author} {\bibfnamefont {L.}~\bibnamefont {Dell'Anna}}, \ and\
  \bibinfo {author} {\bibfnamefont {S.}~\bibnamefont {Pilati}},\ }\bibfield
  {title} {\enquote {\bibinfo {title} {{Projective quantum Monte Carlo
  simulations guided by unrestricted neural network states}},}\ }\href@noop {}
  {\bibfield  {journal} {\bibinfo  {journal} {Phys. Rev. B}\ }\textbf {\bibinfo
  {volume} {98}},\ \bibinfo {pages} {235145} (\bibinfo {year}
  {2018}{\natexlab{b}})}\BibitemShut {NoStop}%
\bibitem [{\citenamefont {Reatto}\ and\ \citenamefont
  {Masserini}(1988)}]{reatto1988shadow}%
  \BibitemOpen
  \bibfield  {author} {\bibinfo {author} {\bibfnamefont {L.}~\bibnamefont
  {Reatto}}\ and\ \bibinfo {author} {\bibfnamefont {G.}~\bibnamefont
  {Masserini}},\ }\bibfield  {title} {\enquote {\bibinfo {title} {{Shadow wave
  function for many-boson systems}},}\ }\href {\doibase
  10.1103/PhysRevB.38.4516} {\bibfield  {journal} {\bibinfo  {journal} {Phys.
  Rev. B}\ }\textbf {\bibinfo {volume} {38}},\ \bibinfo {pages} {4516--4522}
  (\bibinfo {year} {1988})}\BibitemShut {NoStop}%
\bibitem [{\citenamefont {Vitiello}\ \emph {et~al.}(1988)\citenamefont
  {Vitiello}, \citenamefont {Runge},\ and\ \citenamefont
  {Kalos}}]{shadowvitiello}%
  \BibitemOpen
  \bibfield  {author} {\bibinfo {author} {\bibfnamefont {S.}~\bibnamefont
  {Vitiello}}, \bibinfo {author} {\bibfnamefont {K.}~\bibnamefont {Runge}}, \
  and\ \bibinfo {author} {\bibfnamefont {M.~H.}\ \bibnamefont {Kalos}},\
  }\bibfield  {title} {\enquote {\bibinfo {title} {Variational calculations for
  solid and liquid {He} 4 with a shadow wave function},}\ }\href {\doibase
  10.1103/PhysRevLett.60.1970} {\bibfield  {journal} {\bibinfo  {journal}
  {Phys. Rev. Lett.}\ }\textbf {\bibinfo {volume} {60}},\ \bibinfo {pages}
  {1970} (\bibinfo {year} {1988})}\BibitemShut {NoStop}%
\bibitem [{\citenamefont {Deng}\ \emph {et~al.}(2017)\citenamefont {Deng},
  \citenamefont {Li},\ and\ \citenamefont {Das~Sarma}}]{deng2017quantum}%
  \BibitemOpen
  \bibfield  {author} {\bibinfo {author} {\bibfnamefont {D.-L.}\ \bibnamefont
  {Deng}}, \bibinfo {author} {\bibfnamefont {X.}~\bibnamefont {Li}}, \ and\
  \bibinfo {author} {\bibfnamefont {S.}~\bibnamefont {Das~Sarma}},\ }\bibfield
  {title} {\enquote {\bibinfo {title} {Quantum entanglement in neural network
  states},}\ }\href {\doibase 10.1103/PhysRevX.7.021021} {\bibfield  {journal}
  {\bibinfo  {journal} {Phys. Rev. X}\ }\textbf {\bibinfo {volume} {7}},\
  \bibinfo {pages} {021021} (\bibinfo {year} {2017})}\BibitemShut {NoStop}%
\bibitem [{\citenamefont {Gao}\ and\ \citenamefont
  {Duan}(2017)}]{gao2017efficient}%
  \BibitemOpen
  \bibfield  {author} {\bibinfo {author} {\bibfnamefont {X.}~\bibnamefont
  {Gao}}\ and\ \bibinfo {author} {\bibfnamefont {L.-M.}\ \bibnamefont {Duan}},\
  }\bibfield  {title} {\enquote {\bibinfo {title} {Efficient representation of
  quantum many-body states with deep neural networks},}\ }\href {\doibase
  10.1038/s41467-017-00705-2} {\bibfield  {journal} {\bibinfo  {journal} {Nat.
  Commun.}\ }\textbf {\bibinfo {volume} {8}},\ \bibinfo {pages} {662} (\bibinfo
  {year} {2017})}\BibitemShut {NoStop}%
\bibitem [{\citenamefont {Saito}(2017)}]{saito}%
  \BibitemOpen
  \bibfield  {author} {\bibinfo {author} {\bibfnamefont {H.}~\bibnamefont
  {Saito}},\ }\bibfield  {title} {\enquote {\bibinfo {title} {{Solving the
  Bose-Hubbard Model with Machine Learning}},}\ }\href {\doibase
  10.7566/JPSJ.86.093001} {\bibfield  {journal} {\bibinfo  {journal} {J. Phys.
  Soc. Jpn.}\ }\textbf {\bibinfo {volume} {86}},\ \bibinfo {pages} {093001}
  (\bibinfo {year} {2017})}\BibitemShut {NoStop}%
\bibitem [{\citenamefont {Chen}\ \emph {et~al.}(2018)\citenamefont {Chen},
  \citenamefont {Cheng}, \citenamefont {Xie}, \citenamefont {Wang},\ and\
  \citenamefont {Xiang}}]{chen2018equivalence}%
  \BibitemOpen
  \bibfield  {author} {\bibinfo {author} {\bibfnamefont {J.}~\bibnamefont
  {Chen}}, \bibinfo {author} {\bibfnamefont {S.}~\bibnamefont {Cheng}},
  \bibinfo {author} {\bibfnamefont {H.}~\bibnamefont {Xie}}, \bibinfo {author}
  {\bibfnamefont {L.}~\bibnamefont {Wang}}, \ and\ \bibinfo {author}
  {\bibfnamefont {T.}~\bibnamefont {Xiang}},\ }\bibfield  {title} {\enquote
  {\bibinfo {title} {{Equivalence of restricted Boltzmann machines and tensor
  network states}},}\ }\href {\doibase 10.1103/PhysRevB.97.085104} {\bibfield
  {journal} {\bibinfo  {journal} {Phys. Rev. B}\ }\textbf {\bibinfo {volume}
  {97}},\ \bibinfo {pages} {085104} (\bibinfo {year} {2018})}\BibitemShut
  {NoStop}%
\bibitem [{\citenamefont {Glasser}\ \emph {et~al.}(2018)\citenamefont
  {Glasser}, \citenamefont {Pancotti}, \citenamefont {August}, \citenamefont
  {Rodriguez},\ and\ \citenamefont {Cirac}}]{glasser2018neural}%
  \BibitemOpen
  \bibfield  {author} {\bibinfo {author} {\bibfnamefont {I.}~\bibnamefont
  {Glasser}}, \bibinfo {author} {\bibfnamefont {N.}~\bibnamefont {Pancotti}},
  \bibinfo {author} {\bibfnamefont {M.}~\bibnamefont {August}}, \bibinfo
  {author} {\bibfnamefont {I.~D.}\ \bibnamefont {Rodriguez}}, \ and\ \bibinfo
  {author} {\bibfnamefont {J.~I.}\ \bibnamefont {Cirac}},\ }\bibfield  {title}
  {\enquote {\bibinfo {title} {Neural-network quantum states, string-bond
  states, and chiral topological states},}\ }\href {\doibase
  10.1103/PhysRevX.8.011006} {\bibfield  {journal} {\bibinfo  {journal} {Phys.
  Rev. X}\ }\textbf {\bibinfo {volume} {8}},\ \bibinfo {pages} {011006}
  (\bibinfo {year} {2018})}\BibitemShut {NoStop}%
\bibitem [{\citenamefont {Freitas}\ \emph {et~al.}(2018)\citenamefont
  {Freitas}, \citenamefont {Morigi},\ and\ \citenamefont
  {Dunjko}}]{freitas2018neural}%
  \BibitemOpen
  \bibfield  {author} {\bibinfo {author} {\bibfnamefont {N.}~\bibnamefont
  {Freitas}}, \bibinfo {author} {\bibfnamefont {G.}~\bibnamefont {Morigi}}, \
  and\ \bibinfo {author} {\bibfnamefont {V.}~\bibnamefont {Dunjko}},\
  }\bibfield  {title} {\enquote {\bibinfo {title} {{Neural Network Operations
  and Susuki-Trotter evolution of Neural Network States}},}\ }\href
  {https://arxiv.org/abs/1803.02118} {\bibfield  {journal} {\bibinfo  {journal}
  {arXiv preprint arXiv:1803.02118}\ } (\bibinfo {year} {2018})}\BibitemShut
  {NoStop}%
\bibitem [{\citenamefont {Cai}\ and\ \citenamefont
  {Liu}(2018)}]{cai2018approximating}%
  \BibitemOpen
  \bibfield  {author} {\bibinfo {author} {\bibfnamefont {Z.}~\bibnamefont
  {Cai}}\ and\ \bibinfo {author} {\bibfnamefont {J.}~\bibnamefont {Liu}},\
  }\bibfield  {title} {\enquote {\bibinfo {title} {Approximating quantum
  many-body wave functions using artificial neural networks},}\ }\href@noop {}
  {\bibfield  {journal} {\bibinfo  {journal} {Phys. Rev. B}\ }\textbf {\bibinfo
  {volume} {97}},\ \bibinfo {pages} {035116} (\bibinfo {year}
  {2018})}\BibitemShut {NoStop}%
\bibitem [{\citenamefont {Saito}\ and\ \citenamefont {Kato}(2018)}]{saito2}%
  \BibitemOpen
  \bibfield  {author} {\bibinfo {author} {\bibfnamefont {H.}~\bibnamefont
  {Saito}}\ and\ \bibinfo {author} {\bibfnamefont {M.}~\bibnamefont {Kato}},\
  }\bibfield  {title} {\enquote {\bibinfo {title} {Machine learning technique
  to find quantum many-body ground states of bosons on a lattice},}\ }\href
  {\doibase 10.7566/JPSJ.87.014001} {\bibfield  {journal} {\bibinfo  {journal}
  {J. Phys. Soc. Jpn.}\ }\textbf {\bibinfo {volume} {87}},\ \bibinfo {pages}
  {014001} (\bibinfo {year} {2018})}\BibitemShut {NoStop}%
\bibitem [{\citenamefont {Carleo}\ \emph {et~al.}(2018)\citenamefont {Carleo},
  \citenamefont {Nomura},\ and\ \citenamefont
  {Imada}}]{carleo2018constructing}%
  \BibitemOpen
  \bibfield  {author} {\bibinfo {author} {\bibfnamefont {G.}~\bibnamefont
  {Carleo}}, \bibinfo {author} {\bibfnamefont {Y.}~\bibnamefont {Nomura}}, \
  and\ \bibinfo {author} {\bibfnamefont {M.}~\bibnamefont {Imada}},\ }\bibfield
   {title} {\enquote {\bibinfo {title} {Constructing exact representations of
  quantum many-body systems with deep neural networks},}\ }\href@noop {}
  {\bibfield  {journal} {\bibinfo  {journal} {Nat. Commun.}\ }\textbf
  {\bibinfo {volume} {9}},\ \bibinfo {pages} {5322} (\bibinfo {year}
  {2018})}\BibitemShut {NoStop}%
\bibitem [{\citenamefont {Luo}\ and\ \citenamefont
  {Clark}(2018)}]{luo2018backflow}%
  \BibitemOpen
  \bibfield  {author} {\bibinfo {author} {\bibfnamefont {D.}~\bibnamefont
  {Luo}}\ and\ \bibinfo {author} {\bibfnamefont {B.~K.}\ \bibnamefont
  {Clark}},\ }\bibfield  {title} {\enquote {\bibinfo {title} {Backflow
  transformations via neural networks for quantum many-body wave-functions},}\
  }\href@noop {} {\bibfield  {journal} {\bibinfo  {journal} {arXiv preprint
  arXiv:1807.10770}\ } (\bibinfo {year} {2018})}\BibitemShut {NoStop}%
\bibitem [{\citenamefont {Kochkov}\ and\ \citenamefont
  {Clark}(2018)}]{kochkov2018variational}%
  \BibitemOpen
  \bibfield  {author} {\bibinfo {author} {\bibfnamefont {D.}~\bibnamefont
  {Kochkov}}\ and\ \bibinfo {author} {\bibfnamefont {B.~K.}\ \bibnamefont
  {Clark}},\ }\bibfield  {title} {\enquote {\bibinfo {title} {Variational
  optimization in the {AI} era: Computational graph states and supervised
  wave-function optimization},}\ }\href@noop {} {\bibfield  {journal} {\bibinfo
   {journal} {arXiv preprint arXiv:1811.12423}\ } (\bibinfo {year}
  {2018})}\BibitemShut {NoStop}%
\bibitem [{\citenamefont {Ruggeri}\ \emph {et~al.}(2018)\citenamefont
  {Ruggeri}, \citenamefont {Moroni},\ and\ \citenamefont
  {Holzmann}}]{PhysRevLett.120.205302}%
  \BibitemOpen
  \bibfield  {author} {\bibinfo {author} {\bibfnamefont {M.}~\bibnamefont
  {Ruggeri}}, \bibinfo {author} {\bibfnamefont {S.}~\bibnamefont {Moroni}}, \
  and\ \bibinfo {author} {\bibfnamefont {M.}~\bibnamefont {Holzmann}},\
  }\bibfield  {title} {\enquote {\bibinfo {title} {{Nonlinear Network
  Description for Many-Body Quantum Systems in Continuous Space}},}\ }\href
  {\doibase 10.1103/PhysRevLett.120.205302} {\bibfield  {journal} {\bibinfo
  {journal} {Phys. Rev. Lett.}\ }\textbf {\bibinfo {volume} {120}},\ \bibinfo
  {pages} {205302} (\bibinfo {year} {2018})}\BibitemShut {NoStop}%
\bibitem [{\citenamefont {Beach}\ \emph {et~al.}(2019)\citenamefont {Beach},
  \citenamefont {Melko}, \citenamefont {Grover},\ and\ \citenamefont
  {Hsieh}}]{beach2019making}%
  \BibitemOpen
  \bibfield  {author} {\bibinfo {author} {\bibfnamefont {M.~J.}\ \bibnamefont
  {Beach}}, \bibinfo {author} {\bibfnamefont {R.~G.}\ \bibnamefont {Melko}},
  \bibinfo {author} {\bibfnamefont {T.}~\bibnamefont {Grover}}, \ and\ \bibinfo
  {author} {\bibfnamefont {T.~H.}\ \bibnamefont {Hsieh}},\ }\bibfield  {title}
  {\enquote {\bibinfo {title} {Making {Trotters} sprint: A variational
  imaginary time ansatz for quantum many-body systems},}\ }\href@noop {}
  {\bibfield  {journal} {\bibinfo  {journal} {arXiv preprint arXiv:1904.00019}\
  } (\bibinfo {year} {2019})}\BibitemShut {NoStop}%
\bibitem [{\citenamefont {McBrian}\ \emph {et~al.}(2019)\citenamefont
  {McBrian}, \citenamefont {Carleo},\ and\ \citenamefont
  {Khatami}}]{mcbrian2019ground}%
  \BibitemOpen
  \bibfield  {author} {\bibinfo {author} {\bibfnamefont {K.}~\bibnamefont
  {McBrian}}, \bibinfo {author} {\bibfnamefont {G.}~\bibnamefont {Carleo}}, \
  and\ \bibinfo {author} {\bibfnamefont {E.}~\bibnamefont {Khatami}},\
  }\bibfield  {title} {\enquote {\bibinfo {title} {{Ground state phase diagram
  of the one-dimensional Bose-Hubbard model from restricted Boltzmann
  machines}},}\ }\href@noop {} {\bibfield  {journal} {\bibinfo  {journal}
  {arXiv preprint arXiv:1903.03076}\ } (\bibinfo {year} {2019})}\BibitemShut
  {NoStop}%
\bibitem [{\citenamefont {Collura}\ \emph {et~al.}(2019)\citenamefont
  {Collura}, \citenamefont {Del'Anna}, \citenamefont {Felser},\ and\
  \citenamefont {Montangero}}]{collura2019descriptive}%
  \BibitemOpen
  \bibfield  {author} {\bibinfo {author} {\bibfnamefont {M.}~\bibnamefont
  {Collura}}, \bibinfo {author} {\bibfnamefont {L.}~\bibnamefont {Del'Anna}},
  \bibinfo {author} {\bibfnamefont {T.}~\bibnamefont {Felser}}, \ and\ \bibinfo
  {author} {\bibfnamefont {S.}~\bibnamefont {Montangero}},\ }\bibfield  {title}
  {\enquote {\bibinfo {title} {{On the descriptive power of Neural-Networks as
  constrained Tensor Networks with exponentially large bond dimension}},}\
  }\href@noop {} {\bibfield  {journal} {\bibinfo  {journal} {arXiv preprint
  arXiv:1905.11351}\ } (\bibinfo {year} {2019})}\BibitemShut {NoStop}%
\bibitem [{\citenamefont {Kessler}\ \emph {et~al.}(2019)\citenamefont
  {Kessler}, \citenamefont {Calcavecchia},\ and\ \citenamefont
  {K{\"u}hne}}]{kessler2019artificial}%
  \BibitemOpen
  \bibfield  {author} {\bibinfo {author} {\bibfnamefont {J.}~\bibnamefont
  {Kessler}}, \bibinfo {author} {\bibfnamefont {F.}~\bibnamefont
  {Calcavecchia}}, \ and\ \bibinfo {author} {\bibfnamefont {T.~D.}\
  \bibnamefont {K{\"u}hne}},\ }\bibfield  {title} {\enquote {\bibinfo {title}
  {Artificial neural networks as trial wave functions for quantum monte
  carlo},}\ }\href@noop {} {\bibfield  {journal} {\bibinfo  {journal} {arXiv
  preprint arXiv:1904.10251}\ } (\bibinfo {year} {2019})}\BibitemShut {NoStop}%
\bibitem [{\citenamefont {Nagy}\ and\ \citenamefont
  {Savona}(2019)}]{nagy2019variational}%
  \BibitemOpen
  \bibfield  {author} {\bibinfo {author} {\bibfnamefont {A.}~\bibnamefont
  {Nagy}}\ and\ \bibinfo {author} {\bibfnamefont {V.}~\bibnamefont {Savona}},\
  }\bibfield  {title} {\enquote {\bibinfo {title} {Variational quantum monte
  carlo with neural network ansatz for open quantum systems},}\ }\href@noop {}
  {\bibfield  {journal} {\bibinfo  {journal} {arXiv preprint arXiv:1902.09483}\
  } (\bibinfo {year} {2019})}\BibitemShut {NoStop}%
\bibitem [{\citenamefont {Bauer}\ \emph {et~al.}(2011)\citenamefont {Bauer},
  \citenamefont {Carr}, \citenamefont {Evertz}, \citenamefont {Feiguin},
  \citenamefont {Freire}, \citenamefont {Fuchs}, \citenamefont {Gamper},
  \citenamefont {Gukelberger}, \citenamefont {Gull}, \citenamefont {Guertler}
  \emph {et~al.}}]{bauer2011alps}%
  \BibitemOpen
  \bibfield  {author} {\bibinfo {author} {\bibfnamefont {B.}~\bibnamefont
  {Bauer}}, \bibinfo {author} {\bibfnamefont {L.}~\bibnamefont {Carr}},
  \bibinfo {author} {\bibfnamefont {H.~G.}\ \bibnamefont {Evertz}}, \bibinfo
  {author} {\bibfnamefont {A.}~\bibnamefont {Feiguin}}, \bibinfo {author}
  {\bibfnamefont {J.}~\bibnamefont {Freire}}, \bibinfo {author} {\bibfnamefont
  {S.}~\bibnamefont {Fuchs}}, \bibinfo {author} {\bibfnamefont
  {L.}~\bibnamefont {Gamper}}, \bibinfo {author} {\bibfnamefont
  {J.}~\bibnamefont {Gukelberger}}, \bibinfo {author} {\bibfnamefont
  {E.}~\bibnamefont {Gull}}, \bibinfo {author} {\bibfnamefont {S.}~\bibnamefont
  {Guertler}},  \emph {et~al.},\ }\bibfield  {title} {\enquote {\bibinfo
  {title} {{The ALPS project release 2.0: open source software for strongly
  correlated systems}},}\ }\href@noop {} {\bibfield  {journal} {\bibinfo
  {journal} {J. Stat. Mech.: Theory Exp.}\ }\textbf {\bibinfo {volume}
  {2011}},\ \bibinfo {pages} {P05001} (\bibinfo {year} {2011})}\BibitemShut
  {NoStop}%
\bibitem [{\citenamefont {Torlai}\ \emph {et~al.}(2018)\citenamefont {Torlai},
  \citenamefont {Mazzola}, \citenamefont {Carrasquilla}, \citenamefont
  {Troyer}, \citenamefont {Melko},\ and\ \citenamefont
  {Carleo}}]{torlaitomography}%
  \BibitemOpen
  \bibfield  {author} {\bibinfo {author} {\bibfnamefont {G.}~\bibnamefont
  {Torlai}}, \bibinfo {author} {\bibfnamefont {G.}~\bibnamefont {Mazzola}},
  \bibinfo {author} {\bibfnamefont {J.}~\bibnamefont {Carrasquilla}}, \bibinfo
  {author} {\bibfnamefont {M.}~\bibnamefont {Troyer}}, \bibinfo {author}
  {\bibfnamefont {R.}~\bibnamefont {Melko}}, \ and\ \bibinfo {author}
  {\bibfnamefont {G.}~\bibnamefont {Carleo}},\ }\bibfield  {title} {\enquote
  {\bibinfo {title} {Neural-network quantum state tomography},}\ }\href
  {https://doi.org/10.1038/s41567-018-0048-5} {\bibfield  {journal} {\bibinfo
  {journal} {Nat. Phys.}\ }\textbf {\bibinfo {volume} {14}},\ \bibinfo {pages}
  {447--450} (\bibinfo {year} {2018})}\BibitemShut {NoStop}%
\bibitem [{\citenamefont {Sorella}(1998)}]{PhysRevLett.80.4558}%
  \BibitemOpen
  \bibfield  {author} {\bibinfo {author} {\bibfnamefont {S.}~\bibnamefont
  {Sorella}},\ }\bibfield  {title} {\enquote {\bibinfo {title} {Green function
  {Monte Carlo} with stochastic reconfiguration},}\ }\href {\doibase
  10.1103/PhysRevLett.80.4558} {\bibfield  {journal} {\bibinfo  {journal}
  {Phys. Rev. Lett.}\ }\textbf {\bibinfo {volume} {80}},\ \bibinfo {pages}
  {4558--4561} (\bibinfo {year} {1998})}\BibitemShut {NoStop}%
\bibitem [{\citenamefont {Carleo}\ \emph {et~al.}(2019)\citenamefont {Carleo},
  \citenamefont {Choo}, \citenamefont {Hofmann}, \citenamefont {Smith},
  \citenamefont {Westerhout}, \citenamefont {Alet}, \citenamefont {Davis},
  \citenamefont {Efthymiou}, \citenamefont {Glasser}, \citenamefont {Lin} \emph
  {et~al.}}]{carleo2019netket}%
  \BibitemOpen
  \bibfield  {author} {\bibinfo {author} {\bibfnamefont {G.}~\bibnamefont
  {Carleo}}, \bibinfo {author} {\bibfnamefont {K.}~\bibnamefont {Choo}},
  \bibinfo {author} {\bibfnamefont {D.}~\bibnamefont {Hofmann}}, \bibinfo
  {author} {\bibfnamefont {J.~E.}\ \bibnamefont {Smith}}, \bibinfo {author}
  {\bibfnamefont {T.}~\bibnamefont {Westerhout}}, \bibinfo {author}
  {\bibfnamefont {F.}~\bibnamefont {Alet}}, \bibinfo {author} {\bibfnamefont
  {E.~J.}\ \bibnamefont {Davis}}, \bibinfo {author} {\bibfnamefont
  {S.}~\bibnamefont {Efthymiou}}, \bibinfo {author} {\bibfnamefont
  {I.}~\bibnamefont {Glasser}}, \bibinfo {author} {\bibfnamefont {S.-H.}\
  \bibnamefont {Lin}},  \emph {et~al.},\ }\bibfield  {title} {\enquote
  {\bibinfo {title} {Netket: A machine learning toolkit for many-body quantum
  systems},}\ }\href@noop {} {\bibfield  {journal} {\bibinfo  {journal} {arXiv
  preprint arXiv:1904.00031}\ } (\bibinfo {year} {2019})}\BibitemShut {NoStop}%
\bibitem [{\citenamefont {Becca}\ and\ \citenamefont
  {Sorella}(2017)}]{becca_sorella}%
  \BibitemOpen
  \bibfield  {author} {\bibinfo {author} {\bibfnamefont {F.}~\bibnamefont
  {Becca}}\ and\ \bibinfo {author} {\bibfnamefont {S.}~\bibnamefont
  {Sorella}},\ }\href {\doibase 10.1017/9781316417041} {\emph {\bibinfo {title}
  {Quantum {Monte Carlo} Approaches for Correlated Systems}}}\ (\bibinfo
  {publisher} {Cambridge University Press},\ \bibinfo {year}
  {2017})\BibitemShut {NoStop}%
\bibitem [{\citenamefont {Sorella}\ and\ \citenamefont
  {Capriotti}(2000)}]{SorellaCTGFMC}%
  \BibitemOpen
  \bibfield  {author} {\bibinfo {author} {\bibfnamefont {S.}~\bibnamefont
  {Sorella}}\ and\ \bibinfo {author} {\bibfnamefont {L.}~\bibnamefont
  {Capriotti}},\ }\bibfield  {title} {\enquote {\bibinfo {title} {{Green
  function Monte Carlo with stochastic reconfiguration: An effective remedy for
  the sign problem}},}\ }\href {\doibase 10.1103/PhysRevB.61.2599} {\bibfield
  {journal} {\bibinfo  {journal} {Phys. Rev. B}\ }\textbf {\bibinfo {volume}
  {61}},\ \bibinfo {pages} {2599--2612} (\bibinfo {year} {2000})}\BibitemShut
  {NoStop}%
\bibitem [{Note1()}]{Note1}%
  \BibitemOpen
  \bibinfo {note} {One has to accumulate the weight factors of the updates
  occurring within the time-step $\Delta \tau $. See Refs.~\cite
  {SorellaCTGFMC,becca_sorella,inack3} for more details.}\BibitemShut {Stop}%
\bibitem [{\citenamefont {Thijssen}(2007)}]{thijssen}%
  \BibitemOpen
  \bibfield  {author} {\bibinfo {author} {\bibfnamefont {J.}~\bibnamefont
  {Thijssen}},\ }\href {\doibase doi:10.1017/CBO9781139171397} {\emph {\bibinfo
  {title} {Computational physics}}}\ (\bibinfo  {publisher} {Cambridge
  University Press},\ \bibinfo {year} {2007})\BibitemShut {NoStop}%
\bibitem [{\citenamefont {Hetherington}(1984)}]{hetherington1984observations}%
  \BibitemOpen
  \bibfield  {author} {\bibinfo {author} {\bibfnamefont {J.}~\bibnamefont
  {Hetherington}},\ }\bibfield  {title} {\enquote {\bibinfo {title}
  {Observations on the statistical iteration of matrices},}\ }\href@noop {}
  {\bibfield  {journal} {\bibinfo  {journal} {Phys. Rev. A}\ }\textbf {\bibinfo
  {volume} {30}},\ \bibinfo {pages} {2713} (\bibinfo {year}
  {1984})}\BibitemShut {NoStop}%
\bibitem [{\citenamefont {Golinelli}\ \emph {et~al.}(1992)\citenamefont
  {Golinelli}, \citenamefont {Jolicoeur},\ and\ \citenamefont
  {Lacaze}}]{golinelli1992haldane}%
  \BibitemOpen
  \bibfield  {author} {\bibinfo {author} {\bibfnamefont {O.}~\bibnamefont
  {Golinelli}}, \bibinfo {author} {\bibfnamefont {T.}~\bibnamefont
  {Jolicoeur}}, \ and\ \bibinfo {author} {\bibfnamefont {R.}~\bibnamefont
  {Lacaze}},\ }\bibfield  {title} {\enquote {\bibinfo {title} {Haldane gaps in
  a spin-1 heisenberg chain with easy-plane single-ion anisotropy},}\
  }\href@noop {} {\bibfield  {journal} {\bibinfo  {journal} {Phys. Rev. B}\
  }\textbf {\bibinfo {volume} {45}},\ \bibinfo {pages} {9798} (\bibinfo {year}
  {1992})}\BibitemShut {NoStop}%
\bibitem [{\citenamefont {Pollet}\ \emph {et~al.}(2018)\citenamefont {Pollet},
  \citenamefont {Prokof'ev},\ and\ \citenamefont
  {Svistunov}}]{pollet2018stochastic}%
  \BibitemOpen
  \bibfield  {author} {\bibinfo {author} {\bibfnamefont {L.}~\bibnamefont
  {Pollet}}, \bibinfo {author} {\bibfnamefont {N.~V.}\ \bibnamefont
  {Prokof'ev}}, \ and\ \bibinfo {author} {\bibfnamefont {B.~V.}\ \bibnamefont
  {Svistunov}},\ }\bibfield  {title} {\enquote {\bibinfo {title} {Stochastic
  lists: Sampling multivariable functions with population methods},}\ }\href
  {\doibase 10.1103/PhysRevB.98.085102} {\bibfield  {journal} {\bibinfo
  {journal} {Phys. Rev. B}\ }\textbf {\bibinfo {volume} {98}},\ \bibinfo
  {pages} {085102} (\bibinfo {year} {2018})}\BibitemShut {NoStop}%
\bibitem [{Note2()}]{Note2}%
  \BibitemOpen
  \bibinfo {note} {In one and in two dimensional models with short-range
  interactions, the unrestricted Boltzmann machine ansatzes can be mapped to
  constrained matrix product states and to constrained tensor network states,
  respectively~\cite {collura2019descriptive}. In one dimension this mapping
  allows one to implement alternative variational minimization techniques,
  which avoid the sampling over hidden spins.}\BibitemShut {Stop}%
\bibitem [{\citenamefont {Vitiello}\ and\ \citenamefont
  {Whitlock}(1991)}]{whitlock91}%
  \BibitemOpen
  \bibfield  {author} {\bibinfo {author} {\bibfnamefont {S.~A.}\ \bibnamefont
  {Vitiello}}\ and\ \bibinfo {author} {\bibfnamefont {P.~A.}\ \bibnamefont
  {Whitlock}},\ }\bibfield  {title} {\enquote {\bibinfo {title}
  {{Green's-function Monte Carlo algorithm for the solution of the
  Schr\"odinger equation with the shadow wave function}},}\ }\href {\doibase
  10.1103/PhysRevB.44.7373} {\bibfield  {journal} {\bibinfo  {journal} {Phys.
  Rev. B}\ }\textbf {\bibinfo {volume} {44}},\ \bibinfo {pages} {7373--7377}
  (\bibinfo {year} {1991})}\BibitemShut {NoStop}%
\bibitem [{\citenamefont {Sohn}\ and\ \citenamefont
  {Lee}(2012)}]{sohn2012learning}%
  \BibitemOpen
  \bibfield  {author} {\bibinfo {author} {\bibfnamefont {K.}~\bibnamefont
  {Sohn}}\ and\ \bibinfo {author} {\bibfnamefont {H.}~\bibnamefont {Lee}},\
  }\bibfield  {title} {\enquote {\bibinfo {title} {{Learning invariant
  representations with local transformations}},}\ }in\ \href@noop {} {\emph
  {\bibinfo {booktitle} {Proceedings of the 29th International Coference on
  International Conference on Machine Learning}}}\ (\bibinfo {organization}
  {Omnipress},\ \bibinfo {year} {2012})\ pp.\ \bibinfo {pages}
  {1339--1346}\BibitemShut {NoStop}%
\bibitem [{\citenamefont {Choo}\ \emph {et~al.}(2019)\citenamefont {Choo},
  \citenamefont {Neupert},\ and\ \citenamefont {Carleo}}]{choo2019study}%
  \BibitemOpen
  \bibfield  {author} {\bibinfo {author} {\bibfnamefont {K.}~\bibnamefont
  {Choo}}, \bibinfo {author} {\bibfnamefont {T.}~\bibnamefont {Neupert}}, \
  and\ \bibinfo {author} {\bibfnamefont {G.}~\bibnamefont {Carleo}},\
  }\bibfield  {title} {\enquote {\bibinfo {title} {Study of the two-dimensional
  frustrated {J1-J2} model with neural network quantum states},}\ }\href@noop
  {} {\bibfield  {journal} {\bibinfo  {journal} {arXiv preprint
  arXiv:1903.06713}\ } (\bibinfo {year} {2019})}\BibitemShut {NoStop}%
\bibitem [{\citenamefont {{Ferrari}}\ \emph {et~al.}(2019)\citenamefont
  {{Ferrari}}, \citenamefont {{Becca}},\ and\ \citenamefont
  {{Carrasquilla}}}]{2019arXiv190600463F}%
  \BibitemOpen
  \bibfield  {author} {\bibinfo {author} {\bibfnamefont {F.}~\bibnamefont
  {{Ferrari}}}, \bibinfo {author} {\bibfnamefont {F.}~\bibnamefont {{Becca}}},
  \ and\ \bibinfo {author} {\bibfnamefont {J.}~\bibnamefont {{Carrasquilla}}},\
  }\bibfield  {title} {\enquote {\bibinfo {title} {{Neural Gutzwiller-projected
  variational wave functions}},}\ }\href@noop {} {\bibfield  {journal}
  {\bibinfo  {journal} {arXiv preprint arXiv:1906.00463}\ } (\bibinfo {year}
  {2019})}\BibitemShut {NoStop}%
\bibitem [{\citenamefont {Ackley}\ \emph {et~al.}(1985)\citenamefont {Ackley},
  \citenamefont {Hinton},\ and\ \citenamefont
  {Sejnowski}}]{ackley1985learning}%
  \BibitemOpen
  \bibfield  {author} {\bibinfo {author} {\bibfnamefont {D.~H.}\ \bibnamefont
  {Ackley}}, \bibinfo {author} {\bibfnamefont {G.~E.}\ \bibnamefont {Hinton}},
  \ and\ \bibinfo {author} {\bibfnamefont {T.~J.}\ \bibnamefont {Sejnowski}},\
  }\bibfield  {title} {\enquote {\bibinfo {title} {{A learning algorithm for
  Boltzmann machines}},}\ }\href@noop {} {\bibfield  {journal} {\bibinfo
  {journal} {Cogn. Sci.}\ }\textbf {\bibinfo {volume} {9}},\ \bibinfo {pages}
  {147--169} (\bibinfo {year} {1985})}\BibitemShut {NoStop}%
\bibitem [{\citenamefont {Fischer}\ and\ \citenamefont
  {Igel}(2012)}]{fischer2012introduction}%
  \BibitemOpen
  \bibfield  {author} {\bibinfo {author} {\bibfnamefont {A.}~\bibnamefont
  {Fischer}}\ and\ \bibinfo {author} {\bibfnamefont {C.}~\bibnamefont {Igel}},\
  }\bibfield  {title} {\enquote {\bibinfo {title} {{An introduction to
  restricted Boltzmann machines}},}\ }in\ \href@noop {} {\emph {\bibinfo
  {booktitle} {Iberoamerican congress on pattern recognition}}}\ (\bibinfo
  {organization} {Springer},\ \bibinfo {year} {2012})\ pp.\ \bibinfo {pages}
  {14--36}\BibitemShut {NoStop}%
\bibitem [{\citenamefont {Duchi}\ \emph {et~al.}(2011)\citenamefont {Duchi},
  \citenamefont {Hazan},\ and\ \citenamefont {Singer}}]{duchi2011adaptive}%
  \BibitemOpen
  \bibfield  {author} {\bibinfo {author} {\bibfnamefont {J.}~\bibnamefont
  {Duchi}}, \bibinfo {author} {\bibfnamefont {E.}~\bibnamefont {Hazan}}, \ and\
  \bibinfo {author} {\bibfnamefont {Y.}~\bibnamefont {Singer}},\ }\bibfield
  {title} {\enquote {\bibinfo {title} {Adaptive subgradient methods for online
  learning and stochastic optimization},}\ }\href@noop {} {\bibfield  {journal}
  {\bibinfo  {journal} {J. Mach. Learn. Res.}\ }\textbf {\bibinfo {volume}
  {12}},\ \bibinfo {pages} {2121--2159} (\bibinfo {year} {2011})}\BibitemShut
  {NoStop}%
\bibitem [{\citenamefont {Kingma}\ and\ \citenamefont
  {Ba}(2014)}]{kingma2014adam}%
  \BibitemOpen
  \bibfield  {author} {\bibinfo {author} {\bibfnamefont {D.~P.}\ \bibnamefont
  {Kingma}}\ and\ \bibinfo {author} {\bibfnamefont {J.}~\bibnamefont {Ba}},\
  }\bibfield  {title} {\enquote {\bibinfo {title} {Adam: A method for
  stochastic optimization},}\ }\href@noop {} {\bibfield  {journal} {\bibinfo
  {journal} {arXiv preprint arXiv:1412.6980}\ } (\bibinfo {year}
  {2014})}\BibitemShut {NoStop}%
\bibitem [{\citenamefont {Hinton}(2002)}]{hintonConDiv}%
  \BibitemOpen
  \bibfield  {author} {\bibinfo {author} {\bibfnamefont {G.~E.}\ \bibnamefont
  {Hinton}},\ }\bibfield  {title} {\enquote {\bibinfo {title} {Training
  products of experts by minimizing contrastive divergence},}\ }\href {\doibase
  10.1162/089976602760128018} {\bibfield  {journal} {\bibinfo  {journal}
  {Neural Computation}\ }\textbf {\bibinfo {volume} {14}},\ \bibinfo {pages}
  {1771--1800} (\bibinfo {year} {2002})}\BibitemShut {NoStop}%
\bibitem [{\citenamefont {Bengio}\ and\ \citenamefont
  {Delalleau}(2009)}]{bengio2009justifying}%
  \BibitemOpen
  \bibfield  {author} {\bibinfo {author} {\bibfnamefont {Y.}~\bibnamefont
  {Bengio}}\ and\ \bibinfo {author} {\bibfnamefont {O.}~\bibnamefont
  {Delalleau}},\ }\bibfield  {title} {\enquote {\bibinfo {title} {Justifying
  and generalizing contrastive divergence},}\ }\href@noop {} {\bibfield
  {journal} {\bibinfo  {journal} {Neural Comput.}\ }\textbf {\bibinfo {volume}
  {21}},\ \bibinfo {pages} {1601--1621} (\bibinfo {year} {2009})}\BibitemShut
  {NoStop}%
\bibitem [{\citenamefont {{Torlai}}\ and\ \citenamefont
  {{Melko}}(2019)}]{torlai2019review}%
  \BibitemOpen
  \bibfield  {author} {\bibinfo {author} {\bibfnamefont {G.}~\bibnamefont
  {{Torlai}}}\ and\ \bibinfo {author} {\bibfnamefont {R.~G.}\ \bibnamefont
  {{Melko}}},\ }\bibfield  {title} {\enquote {\bibinfo {title} {{Machine
  learning quantum states in the NISQ era}},}\ }\href@noop {} {\bibfield
  {journal} {\bibinfo  {journal} {arXiv e-prints}\ ,\ \bibinfo {pages}
  {arXiv:1905.04312}} (\bibinfo {year} {2019})},\ \Eprint
  {http://arxiv.org/abs/1905.04312} {arXiv:1905.04312 [quant-ph]} \BibitemShut
  {NoStop}%
\bibitem [{\citenamefont {Cabrera}\ and\ \citenamefont
  {Jullien}(1987)}]{cabrera1987role}%
  \BibitemOpen
  \bibfield  {author} {\bibinfo {author} {\bibfnamefont {G.}~\bibnamefont
  {Cabrera}}\ and\ \bibinfo {author} {\bibfnamefont {R.}~\bibnamefont
  {Jullien}},\ }\bibfield  {title} {\enquote {\bibinfo {title} {{Role of
  boundary conditions in the finite-size Ising model}},}\ }\href@noop {}
  {\bibfield  {journal} {\bibinfo  {journal} {Phys. Rev. B}\ }\textbf {\bibinfo
  {volume} {35}},\ \bibinfo {pages} {7062} (\bibinfo {year}
  {1987})}\BibitemShut {NoStop}%
\bibitem [{\citenamefont {Todo}\ and\ \citenamefont
  {Kato}(2001)}]{todo2001cluster}%
  \BibitemOpen
  \bibfield  {author} {\bibinfo {author} {\bibfnamefont {S.}~\bibnamefont
  {Todo}}\ and\ \bibinfo {author} {\bibfnamefont {K.}~\bibnamefont {Kato}},\
  }\bibfield  {title} {\enquote {\bibinfo {title} {{Cluster algorithms for
  general-S quantum spin systems}},}\ }\href@noop {} {\bibfield  {journal}
  {\bibinfo  {journal} {Phys. Rev. Lett.}\ }\textbf {\bibinfo {volume} {87}},\
  \bibinfo {pages} {047203} (\bibinfo {year} {2001})}\BibitemShut {NoStop}%
\bibitem [{\citenamefont {Huang}\ and\ \citenamefont
  {Wang}(2017)}]{huang2017accelerated}%
  \BibitemOpen
  \bibfield  {author} {\bibinfo {author} {\bibfnamefont {L.}~\bibnamefont
  {Huang}}\ and\ \bibinfo {author} {\bibfnamefont {L.}~\bibnamefont {Wang}},\
  }\bibfield  {title} {\enquote {\bibinfo {title} {{Accelerated Monte Carlo
  simulations with restricted Boltzmann machines}},}\ }\href@noop {} {\bibfield
   {journal} {\bibinfo  {journal} {Phys. Rev. B}\ }\textbf {\bibinfo {volume}
  {95}},\ \bibinfo {pages} {035105} (\bibinfo {year} {2017})}\BibitemShut
  {NoStop}%
\bibitem [{\citenamefont {Liu}\ \emph {et~al.}(2017{\natexlab{a}})\citenamefont
  {Liu}, \citenamefont {Shen}, \citenamefont {Qi}, \citenamefont {Meng},\ and\
  \citenamefont {Fu}}]{PhysRevB.95.241104}%
  \BibitemOpen
  \bibfield  {author} {\bibinfo {author} {\bibfnamefont {J.}~\bibnamefont
  {Liu}}, \bibinfo {author} {\bibfnamefont {H.}~\bibnamefont {Shen}}, \bibinfo
  {author} {\bibfnamefont {Y.}~\bibnamefont {Qi}}, \bibinfo {author}
  {\bibfnamefont {Z.~Y.}\ \bibnamefont {Meng}}, \ and\ \bibinfo {author}
  {\bibfnamefont {L.}~\bibnamefont {Fu}},\ }\bibfield  {title} {\enquote
  {\bibinfo {title} {{Self-learning Monte Carlo method and cumulative update in
  fermion systems}},}\ }\href {\doibase 10.1103/PhysRevB.95.241104} {\bibfield
  {journal} {\bibinfo  {journal} {Phys. Rev. B}\ }\textbf {\bibinfo {volume}
  {95}},\ \bibinfo {pages} {241104} (\bibinfo {year}
  {2017}{\natexlab{a}})}\BibitemShut {NoStop}%
\bibitem [{\citenamefont {Liu}\ \emph {et~al.}(2017{\natexlab{b}})\citenamefont
  {Liu}, \citenamefont {Qi}, \citenamefont {Meng},\ and\ \citenamefont
  {Fu}}]{PhysRevB.95.041101}%
  \BibitemOpen
  \bibfield  {author} {\bibinfo {author} {\bibfnamefont {J.}~\bibnamefont
  {Liu}}, \bibinfo {author} {\bibfnamefont {Y.}~\bibnamefont {Qi}}, \bibinfo
  {author} {\bibfnamefont {Z.~Y.}\ \bibnamefont {Meng}}, \ and\ \bibinfo
  {author} {\bibfnamefont {L.}~\bibnamefont {Fu}},\ }\bibfield  {title}
  {\enquote {\bibinfo {title} {{Self-learning Monte Carlo method}},}\ }\href
  {\doibase 10.1103/PhysRevB.95.041101} {\bibfield  {journal} {\bibinfo
  {journal} {Phys. Rev. B}\ }\textbf {\bibinfo {volume} {95}},\ \bibinfo
  {pages} {041101} (\bibinfo {year} {2017}{\natexlab{b}})}\BibitemShut
  {NoStop}%
\bibitem [{\citenamefont {Shen}\ \emph {et~al.}(2018)\citenamefont {Shen},
  \citenamefont {Liu},\ and\ \citenamefont {Fu}}]{PhysRevB.97.205140}%
  \BibitemOpen
  \bibfield  {author} {\bibinfo {author} {\bibfnamefont {H.}~\bibnamefont
  {Shen}}, \bibinfo {author} {\bibfnamefont {J.}~\bibnamefont {Liu}}, \ and\
  \bibinfo {author} {\bibfnamefont {L.}~\bibnamefont {Fu}},\ }\bibfield
  {title} {\enquote {\bibinfo {title} {{Self-learning Monte Carlo with deep
  neural networks}},}\ }\href {\doibase 10.1103/PhysRevB.97.205140} {\bibfield
  {journal} {\bibinfo  {journal} {Phys. Rev. B}\ }\textbf {\bibinfo {volume}
  {97}},\ \bibinfo {pages} {205140} (\bibinfo {year} {2018})}\BibitemShut
  {NoStop}%
\bibitem [{\citenamefont {Li}\ \emph {et~al.}(2019)\citenamefont {Li},
  \citenamefont {Dee}, \citenamefont {Khatami},\ and\ \citenamefont
  {Johnston}}]{li2019accelerating}%
  \BibitemOpen
  \bibfield  {author} {\bibinfo {author} {\bibfnamefont {S.}~\bibnamefont
  {Li}}, \bibinfo {author} {\bibfnamefont {P.~M.}\ \bibnamefont {Dee}},
  \bibinfo {author} {\bibfnamefont {E.}~\bibnamefont {Khatami}}, \ and\
  \bibinfo {author} {\bibfnamefont {S.}~\bibnamefont {Johnston}},\ }\bibfield
  {title} {\enquote {\bibinfo {title} {{Accelerating lattice quantum Monte
  Carlo simulation using artificial neural networks: an application to the
  Holstein model}},}\ }\href@noop {} {\bibfield  {journal} {\bibinfo  {journal}
  {arXiv preprint arXiv:1905.07440}\ } (\bibinfo {year} {2019})}\BibitemShut
  {NoStop}%
\end{thebibliography}

%

\end{document}